\documentclass[reprint,superscriptaddress,aps,prx]{revtex4-2}
\usepackage[T1]{fontenc}
\usepackage[latin1]{inputenc}
\usepackage{lmodern}
\expandafter\let\csname equation*\endcsname\relax
\expandafter\let\csname endequation*\endcsname\relax
\usepackage{amsmath}
\usepackage{amssymb,dsfont}
\usepackage{graphicx}
\usepackage{xcolor}
\usepackage{natbib}
\usepackage{bm}
\usepackage{bbold}
\usepackage[mathscr]{euscript}
\usepackage[cal=boondoxo]{mathalfa}
\usepackage{comment}
\renewcommand\footnotemark{}

\usepackage{braket}
\usepackage[normalem]{ulem}
\usepackage{lineno}

\def\ii{{\rm i}}  
\def\GG{{\bf G}}

\def\db{\boldsymbol{\wp}} 
  
\def\rb{{\bf r}}

\def\hats{\hat{\sigma}}

\def\bra#1{\mathinner{\langle{#1}|}}
\def\ket#1{\mathinner{|{#1}\rangle}}
\def\braket#1{\mathinner{\langle{#1}\rangle}}
\def\jop{\hat{\mathcal{O}}}
\def\dop{\hat{\mathcal{D}}}

\newcommand\varpm{\mathbin{\vcenter{\hbox{%
  \oalign{\hfil$\scriptstyle+$\hfil\cr
          \noalign{\kern-.3ex}
          $\scriptscriptstyle({-})$\cr}%
}}}}

\def\Rdet{\hat{R}_{\mathrm{det.}}}

\begin{document}

\title{Dicke superradiance in ordered arrays of multilevel atoms}
\author{Stuart J. Masson}
\email{s.j.masson@columbia.edu}
\affiliation{Department of Physics, Columbia University, New York, NY 10027, USA}
\author{Jacob P. Covey}
\affiliation{Department of Physics, The University of Illinois at Urbana-Champaign, Urbana, IL 61801, USA}
\author{Sebastian Will}
\affiliation{Department of Physics, Columbia University, New York, NY 10027, USA}
\author{Ana Asenjo-Garcia}
\email{ana.asenjo@columbia.edu}
\affiliation{Department of Physics, Columbia University, New York, NY 10027, USA}

\date{\today}

\begin{abstract}
In inverted atomic ensembles, photon-mediated interactions give rise to Dicke superradiance, a form of many-body decay that results in a rapid release of energy as a photon burst. While originally studied in pointlike ensembles, this phenomenon persists in extended ordered systems if the inter-particle distance is below a certain bound. Here, we investigate Dicke superradiance in a realistic experimental setting using ordered arrays of alkaline-earth(-like) atoms, such as strontium and ytterbium. Such atoms offer exciting new opportunities for light-matter interactions as their internal structure allows for trapping at short interatomic distances compared to their long-wavelength transitions, providing the potential for collectively enhanced dissipative interactions. Despite their intricate electronic structure, we show that two-dimensional arrays of these atomic species should exhibit many-body superradiance for achievable lattice constants. Moreover, superradiance effectively ``closes'' transitions, such that multilevel atoms become more two-level like. This occurs because the avalanchelike decay funnels the emission of most photons into the dominant transition, overcoming the single-atom decay ratios dictated by their fine structure and Zeeman branching. Our work represents an important step in harnessing alkaline-earth atoms as quantum optical sources and as platforms to explore many-body dissipative dynamics.
\end{abstract}

\maketitle

\section{Introduction}

\begin{figure*}[t!]
\begin{center}
\includegraphics[width=0.95\textwidth]{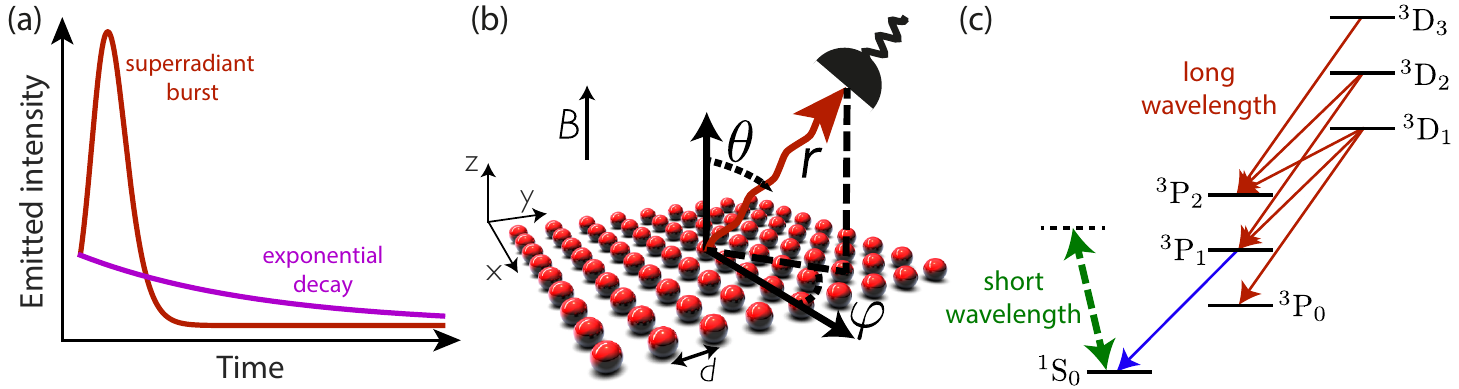}
\caption{(a) Atoms at a point emit a superradiant burst, with a peak emission rate that scales as the square of the number of atoms, in contrast to uncorrelated atoms, which emit an exponentially-decaying pulse. (b) The schematics of the proposed setup: a two-dimensional (2D) array of $N$ atoms with lattice constant $d$ is held in the $x-y$ plane, with the quantization axis set by a magnetic field along the $z$-axis. Light is measured in the far field, at a location described by spherical coordinates $\set{r,\theta,\varphi}$, where $r \gg \sqrt{N}d$. (c)~The relevant level structure of bosonic AEAs. The atoms are optically trapped via strong transitions at short wavelengths (dashed line). The atoms are then prepared in a $^3$D$_{J}$ state, where they decay to $^3$P$_{J}$ states emitting light with a (relatively long) infrared wavelength, and then potentially decay further to the $^1$S$_0$ state. Possible decay paths are indicated by the solid lines. Due to the difference in wavelengths, the decay dynamics from $^3$D$_{J}$ will be dictated by many-body effects.
\label{Fig1}}
\end{center}
\end{figure*}

Atoms in a cavity emit into the same electromagnetic mode, leading to interactions between them, and a collective interaction between light and matter. Interactions are well understood within the paradigm of cavity quantum electrodynamics (QED), as the indistinguishability of the atoms enables their description as a large spin coupled to a single radiative channel. An emblematic example of many-body physics in cavity QED is Dicke superradiance~\cite{Dicke54,Gross82,BenedictBook,Raimond82,Slama07}, where macroscopically verted atoms decay by radiating light in a short bright pulse with peak emission rate that scales quadratically with atom number [see Fig.~\ref{Fig1}(a)]. Dicke superradiance has also been observed in Bose-Einstein condensates~\cite{Inouye99,Schneble03,Yoshikawa05}, where a macroscopically occupied state couples to light. In these scenarios, superradiance is well understood because the permutational symmetry arising from indistinguishability restricts dynamics to a small subspace of the full Hilbert space.

An understanding of collective light-matter interactions beyond the cavity QED regime is critical not only from a fundamental point of view, but also to realize applications in quantum nonlinear optics, quantum simulation, and metrology. Potentially, one could translate concepts such as the superradiant laser~\cite{Meiser09,Bohnet12}, driven-dissipative phase transitions~\cite{Baumann10,Zhiqiang18}, and quantum-enhanced sensing~\cite{Hosten16,Cox16,Colombo22} into a much larger class of systems. For instance, atomic arrays in the single-excitation regime have been proposed as promising platforms for generating novel light sources and optical components, with the recent realization of an atomically-thin mirror~\cite{Bettles16PRL,Shahmoon17,Manzoni18,Rui20,Srakaew23} as an example. The many-body landscape offers a far greater toolbox, and could open up possibilities to create sources of light with unusual statistical properties~\cite{Clemens03,Scully06,Bhatti15,Gulfam18,Liberal19,Holzinger21} or to generate entangled atomic states via dissipation~\cite{Guerin16,Solano17,Hebenstreit17,Masson20PRL,Cipris21,Ferioli21PRX,PineiroOrioli22,Santos22,Rubies23PRA}.

In extended systems in free space, interactions between atoms depend on their relative positions. Theoretical studies of Dicke superradiance in this regime have been greatly limited, as the broken permutational symmetry increases the complexity of the problem, which in principle scales exponentially with the atom number. However, experiments have confirmed that superradiant bursts can still occur. The first demonstrations were performed with thermal molecular and atomic vapors~\cite{Skribanowitz73,Flusberg76,Gross76,Vrehen77,Gounand79}, but observations have since been made in several other systems~\cite{Meinardi03,Scheibner07,Raino18,Ferioli21PRL}.

Ordered atomic arrays~\cite{Bakr10,Sherson10,Kim16,Endres16,Barredo16,Kumar18} have been recently suggested as a promising platform to study many-body decay~\cite{Masson20PRL,Masson22,Sierra22,Robicheaux21PRA_Directional,Rubies22PRA}. In contrast to other setups that typically suffer from dephasing arising from thermal motion or coherent (i.e., Hamiltonian) dipole-dipole interactions, atomic arrays are supposed to experience less dephasing, as the role of Hamiltonian dipole-dipole interactions in the burst is significantly reduced due to the spatial order. In these systems, atoms can decay into many radiative channels. Nevertheless, it has been shown that signatures of superradiance should persist in very extended two-dimensional (2D) systems, of size much larger than the transition wavelength~\cite{Robicheaux21PRA_Directional,Masson22,Sierra22}.

Here, we propose the use of alkaline-earth(-like) atoms (AEAs) in atomic arrays to observe and control Dicke superradiance. These atoms have favorable transitions that enable their trapping at relatively small distances in comparison to the wavelength of the emitted photons. While the atoms are intrinsically multilevel in nature, we demonstrate that the internal competition presented by the different transitions does not prohibit Dicke superradiance. Via a cumulant expansion, we approximate the dynamics and compute the superradiant bursts that would be emitted by arrays of lattice constants that can be achieved in state-of-the-art experimental setups. The emitted light is nontrivially dependent on the geometry of the array and detector location [as shown in Fig.~\ref{Fig1}(b)]. For example, as the interatomic distance is increased, the superradiant burst is lost but then reappears. We show that this dependence can be easily predicted by the use of conditional two-photon correlation functions. Finally, we show how to use Dicke superradiance to inhibit or enhance decay into a particular state, overcoming limits set by fine structure and Zeeman branching.

Our results show that AEAs offer significant advantages to explore and harness superradiance in atom arrays. Such a proof-of-concept experiment would take the first step in the exploration of atomic arrays as a platform for many-body quantum optics in general. In contrast to other phenomena (such as subradiance), superradiance is attractive from an experimental point of view, as it is robust under many imperfections and does not require single-photon detection. The fast and transient nature of the superradiant burst is particularly well suited to these types of experiments, as steady-state phenomena, such as driven superradiance, require long experimental times at which the atoms can explore the full complexity of AEA electronic structure.

The paper is structured as follows. In Section~\ref{structure}, we introduce the full relevant structure of AEAs. In Section~\ref{multicavity}, we consider toy models of multilevel atoms at a single spatial location. We show that Dicke superradiance occurs both for decay to multiple ground states and for cascaded decay (i.e., where the excited state decays to intermediate states before decaying to the final ground state). This allows us to simplify the full level structure of AEAs, keeping only relevant transitions. In Section~\ref{methods}, we introduce the methods necessary to treat these simplified AEAs in ordered arrays with finite separation. In Section~\ref{results}, we show that significant bursts can be achieved in reasonably sized arrays of AEAs, and that this decay can be tailored via the lattice constant.

\section{Transitions of alkaline-earth atoms\label{structure}}

Here, we discuss the relevant atomic transitions of AEAs. These bielectron species have different wavelength transitions that, in theory, allow for trapping and cooling on a short wavelength and for realizing quantum optics experiments on a much longer wavelength~\cite{Olmos13} [see Fig.~\ref{Fig1}(c)]. In particular, the $^1$S$_0$ and metastable $^3$P$_{\{0,2\}}$ states can be trapped at an optical wavelength. If the atoms are excited into a state in the $^3$D$_{J}$ manifold, decay occurs at infrared wavelengths, relative to which the atoms have significantly subwavelength spacing. We consider the bosonic isotopes $^{88}$Sr and $^{174}$Yb, where there is no nuclear spin and thus no hyperfine splitting, for the sake of simplicity. The results can be extended to fermionic isotopes, where similar physics should be observable.

The internal structure of AEAs is well characterized due to their excellent performance as optical atomic clocks~\cite{Ludlow08,Hinkley13,Bloom14,Ushijima15,Nemitz16,Schioppo17,Koller17,Madjarov19,Young20}. In recent years, AEA arrays have also attracted much attention as candidates for quantum computing~\cite{Daley08,Gorshkov09,NChen22,Wu22}, with significant advancements with both strontium~\cite{Cooper18,Norcia18PRX,Barnes22} and ytterbium~\cite{Saskin19,Jenkins22,Okuno22,Ma22}. Current tweezer-array implementations use Rydberg states to mediate interactions and do not require subwavelength spacing. Nevertheless, quantum gas microscopes of $^{174}$Yb have been demonstrated, with interatomic spacings of 266~nm~\cite{Miranda15,Yamamoto16}.

\begin{table}
    \centering
    \begin{tabular}{|c|c|c|}\hline
     \textbf{transition} & \textbf{wavelength (nm)} & \textbf{decay rate ($\times 10^6$ s$^{-1}$)} \\\hline
       $^3$P$_1$ $\rightarrow$ $^1$S$_0$  &  556~\cite{Meggers78} & $1$~\cite{Porsev99} \\\hline
       $^3$D$_1$ $\rightarrow$ $^3$P$_0$  &  1389~\cite{Meggers78} & $2$~\cite{Porsev99} \\\hline
       $^3$D$_1$ $\rightarrow$ $^3$P$_1$  &  1540~\cite{Meggers78} & $1$~\cite{Porsev99} \\\hline
       $^3$D$_1$ $\rightarrow$ $^3$P$_2$  &  2090~\cite{Meggers78} & $0.03$~\cite{Porsev99} \\\hline
       $^3$D$_2$ $\rightarrow$ $^3$P$_1$  &  1480~\cite{Meggers78} & $2$~\cite{Porsev99} \\\hline
       $^3$D$_2$ $\rightarrow$ $^3$P$_2$  &  1980~\cite{Meggers78} & $0.3$~\cite{Porsev99} \\\hline
       $^3$D$_3$ $\rightarrow$ $^3$P$_2$  &  1800~\cite{Meggers78} & $2$~\cite{Porsev99} \\\hline
    \end{tabular}
    \caption{Wavelengths and decay rates for relevant transitions in $^{174}$Yb.}
    \label{Table1}
\end{table}

\begin{table}
    \centering
    \begin{tabular}{|c|c|c|}\hline
     \textbf{transition} & \textbf{wavelength (nm)} & \textbf{decay rate ($\times 10^{5}$ s$^{-1}$)} \\\hline
       $^3$P$_1$ $\rightarrow$ $^1$S$_0$  &  689~\cite{Mickelson09} & $0.47$~\cite{Mickelson09} \\\hline
       $^3$D$_1$ $\rightarrow$ $^3$P$_0$  &  2600~\cite{Sansonetti10} & $2.8$~\cite{ZhangS20} \\\hline
       $^3$D$_1$ $\rightarrow$ $^3$P$_1$  &  2740~\cite{Sansonetti10} & $1.8$~\cite{ZhangS20} \\\hline
       $^3$D$_1$ $\rightarrow$ $^3$P$_2$  &  3070~\cite{Sansonetti10} & $0.088$~\cite{ZhangS20} \\\hline
       $^3$D$_2$ $\rightarrow$ $^3$P$_1$  &  2690~\cite{Sansonetti10} & $3.3$~\cite{Mickelson09} \\\hline
       $^3$D$_2$ $\rightarrow$ $^3$P$_2$  &  3010~\cite{Sansonetti10} & $0.79$~\cite{Mickelson09} \\\hline
       $^3$D$_3$ $\rightarrow$ $^3$P$_2$  &  2920~\cite{Sansonetti10} & $5.9$~\cite{Bowden19}\\\hline
    \end{tabular}
    \caption{Wavelengths and decay rates for relevant transitions in $^{88}$Sr.}
    \label{Table2}
\end{table}

$^{174}$Yb can be operated as an optical source at telecom wavelengths, as the $^3$D$_1$ $\rightarrow$ $^3$P$_{\{0,1\}}$ and $^3$D$_2$ $\rightarrow$ $^3$P$_1$ transitions have wavelengths of around $1.4-1.5~\,\mu$m. Therefore, the light emitted on these transitions is compatible with low-loss fiber-optic cables and devices built with these atoms can be integrated into distributed photonic networks without the need for quantum frequency conversion~\cite{Covey19PRAppl,Covey19PRA,Huie21}. Alternatively, two-level systems can be found on the $^3$D$_3$ $\rightarrow$ $^3$P$_2$ line. In addition, lasers and optical components are readily available for all these transitions. Full details of the transition wavelengths and decay rates for ytterbium are given in Table~\ref{Table1}.

In $^{88}$Sr the ratio between trapping and science wavelengths is even more beneficial to realize closely packed arrays. In particular, atoms initialized in the $^3$D$_3$ state decay at a wavelength of 2.9~$\,\mu$m. However, these transitions are in the midinfrared, where sources, detectors, and other components are less readily available. Full details of the transition wavelengths and decay rates for strontium are given in Table~\ref{Table2}.

Longer-wavelength transitions also exist in alkali atoms, including at telecom frequencies~\cite{Fahey11,Moran16,Uphoff16,Menon20}. However, the lack of metastable states means that the required initial state is more difficult to prepare. Intermediate states have significantly larger linewidths, such that the simplifications we make to the level structure for AEAs are not necessarily valid for alkalis. Additionally, the relatively small fine and hyperfine splitting combined with large multiplicity yields a cluttered spectrum.

\section{Multilevel atoms at a point\label{multicavity}}

We first consider a toy model where atoms are all at the same spatial location and are initially in the excited state. This endows the system with enough symmetry that exact dynamics can be calculated for large atom number. It is well established that superradiance can still occur if there are multiple ground states~\cite{Lin12,Sutherland17,PineiroOrioli22}. Here, we show how the properties of the decay change with the atom number, allowing us to simplify the level structure of the considered AEAs in Section~\ref{results}.

Atoms at a point are indistinguishable to the field. Therefore, the dynamics must respect permutational symmetry and the action of the Hamiltonian is limited to a trivial global frequency shift that can be ignored. This setup is equivalent to atoms identically and resonantly coupled to a ``bad'' cavity, where cavity loss dominates over reabsorption of cavity photons. Furthermore, decay is diagonalized into the action of symmetric spin-lowering operators of the form $\hat{S}_{ge} = \sum_{j=1}^N\hat{\sigma}_{ge}^j$
where $\hat{\sigma}_{ge}^j = \ket{g}_j \bra{e}_j$ is the lowering operator between states $\ket{e}$ and $\ket{g}$ for atom $j$. The complexity of this problem scales as $\mathcal{O}(N^{m-1})$ for $m$-level atoms, making use of the permutational symmetry and the conserved total atom number. The photon emission rate on the transition $\ket{e}\rightarrow\ket{g}$ is calculated as
\begin{equation}
R = \Gamma_0^{eg}\braket{\hat{S}_{eg}\hat{S}_{ge}}.
\end{equation}

\subsection{Multiple ground states: $\Lambda$ systems\label{lambda}}

\begin{figure*}
\includegraphics[width=0.95\textwidth]{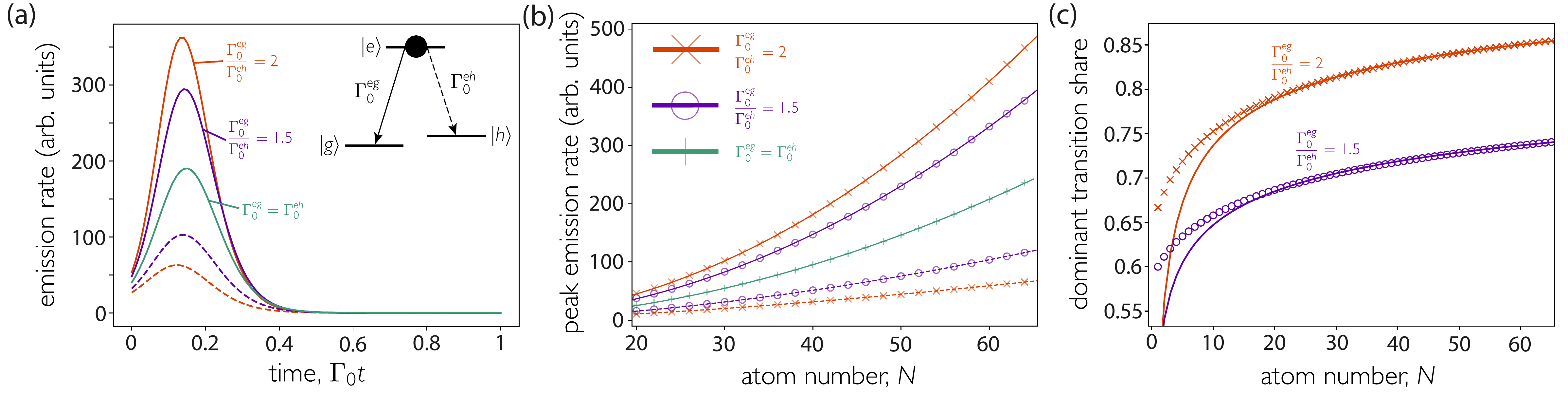}
\caption{Superradiant decay from $\Lambda$ atoms at a point. Each atom decays at a total rate $\Gamma_0 = \Gamma_0^{eg} + \Gamma_0^{eh}$ split between two levels. (a) Superradiant bursts emitted by 40 $\Lambda$ atoms. The solid lines indicate emission on the brighter transition $\ket{e}\rightarrow\ket{g}$, while the dashed lines indicate emission on the less bright transition $\ket{e}\rightarrow\ket{h}$. (b) The scaling of the peak emission on each transition. The solid lines are power-law best fits of data from $N \geq 20$. The solid fit lines indicate a brighter transition, while the dashed fit lines indicate a less bright transition. The scalings are $\sim N^{2.01}$ and $\sim N^{2.00}$ for the brighter transitions with $\Gamma_0^{eg} = 2\Gamma_0^{eh}$ and $\Gamma_0^{eg} = 1.5\Gamma_0^{eh}$, respectively, while the less bright transitions scale as $N^{1.56}$ and $N^{1.72}$. In the balanced case, the scaling is $N^{1.92}$ for both pathways. (c) The fraction of photons emitted on the brighter transition. The solid lines are lines of best fit of data from $N \geq 20$ of the form $1-A/N^B$ with $A=0.541 (0.513)$ and $B=0.31 (0.16)$ for $\Gamma_0^{eg} = 2 (1.5)\Gamma_0^{eh}$, respectively.\label{Fig2}}
\end{figure*}

We now consider a $\Lambda$-system in which the excited state can decay to two different ground states. This toy model is broadly relevant for collective decay in AEAs, where most states can decay to multiple ground states at different frequencies. For our toy model, we assume the frequencies of these transitions to be far separated such that the channels are independent. In the limit of atoms at a point, the dynamics follow the master equation
\begin{equation}
\dot{\rho} = \Gamma_0^{eg}\mathcal{l}[\hat{S}_{ge}](\rho) + \Gamma_0^{eh}\mathcal{l}[\hat{S}_{he}](\rho),
\end{equation}
where decay is diagonalized into collective lowering operators $\hat{S}_{ge,he} = \sum_{j=1}^N \hat{\sigma}^j_{ge,he}$ and
\begin{equation}
\mathcal{l}[\hat{A}](\rho) = \hat{A}\rho\hat{A}^\dagger - \frac{1}{2}\hat{A}^\dagger\hat{A}\rho - \frac{1}{2}\rho\hat{A}^\dagger\hat{A}.
\end{equation}

Superradiant bursts can be emitted on multiple channels at the same time, as shown in Fig.~\ref{Fig2}. Both the height of the burst and its scaling with $N$ depend on the relative strength of the decay channels. For channels of equal decay rate, $\Gamma_0^{eg} = \Gamma_0^{eh}$, the best fit for the scaling of the peaks is $N^{1.92}$, instead of the $N^{2}$ scaling for two-level systems at a point. For imbalanced channels, $\Gamma_0^{eg} > \Gamma_0^{eh}$, the larger burst scales faster than for balanced channels. For the relative rates of decay of $2:1$ and $1.5:1$, the best-fit scalings of the peak photon emission rate on the stronger transition are $N^{2.01}$ and $N^{2.00}$ respectively.  This implies that in such configurations, as long as there is some bias toward one transition, for large enough $N$, the peak on that transition will always scale as the ideal two-level case with $N^2$. In the case of balanced channels, neither channel gains any advantage, and so the scaling is reduced. Furthermore, the superradiant burst on the weaker transition has a peak that scales more slowly than the balanced case. For $\Gamma_0^{eg} = 2\Gamma_0^{eh}$, the scaling is $N^{1.56}$ and for $\Gamma_0^{eg} = 1.5\Gamma_0^{eh}$ it is $N^{1.72}$.

\begin{figure}[t!]
\includegraphics[width=0.4\textwidth]{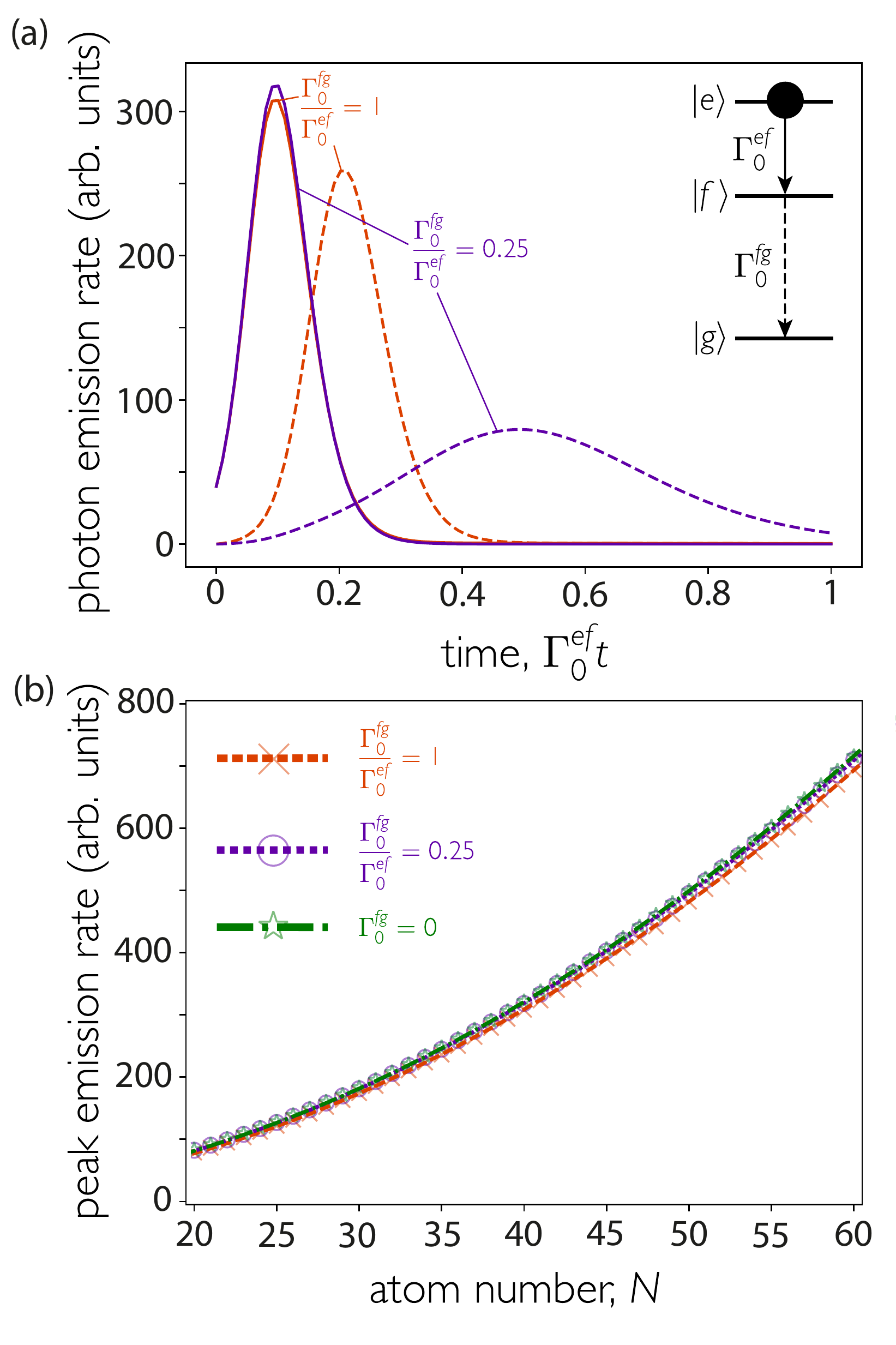}
\caption{Superradiant decay from ladder atoms at a point. Each atom decays initially at a rate $\Gamma_0^{ef}$ to an intermediate state that itself decays at a rate $\Gamma_0^{fg}$. (a) Superradiant bursts emitted by 40 ladder atoms. The solid (dashed) lines indicate emission on the initial (secondary) transition. (b) The scaling of the peak emission by $N$ ladder atoms. The lines are power-law best fits of data from $N \geq 20$. In all three cases the fit scales as approximately $N^2$.\label{Fig3}}
\end{figure}

The percentage of photons emitted on the bright channel increases with the atom number, with the approximation of the light emitted on the weaker transition scaling as a power-law fitting well, as shown in Fig.~\ref{Fig2}(c). Significant population accumulates in $\ket{g}$ faster than in $\ket{h}$ and so the collective enhancement of Dicke superradiance occurs earlier, ``stealing'' photons from the weaker transition. For large atom number, if the ratio between decay rates is strongly biased toward the brighter transition, the impact of the weaker transition is minimal. The bias of the imbalance of photons emitted on each transition has been reported in Ref.~\cite{Sutherland17,PineiroOrioli22}.

\subsection{Cascaded decay: Ladder systems}

We now consider a ladder system in which the excited state, $\ket{e}$, decays to an intermediate state $\ket{f}$, that itself decays again to the ground state $\ket{g}$. This toy model is relevant to collective decay in AEAs from states that decay to the $^3$P$_1$ manifold, as those states can themselves decay to the $^1$S$_0$ state at similar rates to the initial decay. In the limit of all atoms at a point, the dynamics follow the master equation
\begin{equation}
\dot{\rho} = \Gamma_0^{ef}\mathcal{l}[\hat{S}_{ef}](\rho) + \Gamma_0^{fg}\mathcal{l}[\hat{S}_{fg}](\rho),
\end{equation}
where the decay is diagonalized into the collective lowering operators $\hat{S}_{ef,fg} = \sum_{j=1}^N \hat{\sigma}_{ef,fg}^j$.

A superradiant burst is emitted on both transitions consecutively, as shown in Fig.~\ref{Fig3}(a). This occurs because the excited state decay is extremely fast due to large population inversion, while the decay of the intermediate state is very slow due to small inversion. By the time that the population in the intermediate state is large enough to drive fast collective decay, the superradiant burst from the first transition is mostly finished. In the regime $N\Gamma_0^{ef} \gg \Gamma_0^{fg}$, the scaling of the first superradiant peak goes approximately as $\sim N^{2}$ regardless of the relative ratio of decays and the two-level case is retrieved [see Fig.~\ref{Fig3}(b)].

\section{Theoretical methods for ordered extended arrays\label{methods}}

\subsection{Spin model for multilevel atoms}

Here we introduce the theoretical framework to investigate an array of $N$ multilevel atoms that interact with each other via free space beyond the point approximation. Without permutational symmetry, atoms interact both coherently and dissipatively. Under a Born-Markov approximation, the atomic density matrix evolves according to the master equation~\cite{Gruner96,Dung02}
\begin{equation}
\dot{\rho} = \sum\limits_{a} - \frac{\ii}{\hbar} [\mathcal{H}_{a},\rho] + \mathcal{L}_{a}(\rho),\label{fullme}
\end{equation}
where an excited state $\ket{e}$ decays to a set of ground states $\ket{g_a}$. Each Hamiltonian and Lindbladian read
\begin{align}
\mathcal{H}_{a} &=  - \hbar\omega_{a} \sum\limits_{j=1}^N \hats^j_{g_ag_a} + \sum\limits_{j,l=1}^N J_{jl}^{a} \hats_{eg_a}^j \hats_{g_ae}^l, \\
\mathcal{L}_{a}(\rho) &= \sum\limits_{j,l=1}^N \frac{\Gamma_{jl}^{a}}{2} \left(2\hats_{g_ae}^l \rho \hats_{eg_a}^j - \hats_{eg_a}^j \hats_{g_ae}^l\rho - \rho\hats_{eg_a}^j \hats_{g_ae}^l \right)\label{lind},
\end{align}
where $\omega_{a}$ is the frequency of the transition from $\ket{e} \rightarrow \ket{g_a}$, $\hats_{g_ae}^j = \ket{g_a}_j\bra{e}_j$ is the lowering operator from state $\ket{e}\rightarrow\ket{g_a}$ for the $j$th atom, and interactions between atoms $j$ and $l$ are characterized by
\begin{equation}
J_{jl}^{a} - \frac{\ii \Gamma_{jl}^{a}}{2} = -\frac{\mu_0\omega_{a}^2}{\hbar} \db^*_{a} \cdot \GG_0(\rb_j,\rb_l,\omega_{a}) \cdot \db_{a}.
\end{equation}
Here, $\db_{a}$ is the normalized dipole-matrix element of the transition and $\GG_0(\rb_j,\rb_l,\omega_{a})$ is the electromagnetic field propagator between atoms at positions $\rb_j$ and $\rb_l$~\cite{Asenjo17PRA}. The dissipator of Eq.~\eqref{lind} can be expressed in terms of a set of collective lowering operators
\begin{align}
\mathcal{L}_{a}(\rho) = \sum\limits_{\nu=1}^N \frac{\Gamma_{\nu}^{a}}{2} \left(2\jop_{\nu,a} \rho \,\jop_{\nu,a}^\dagger - \jop_{\nu,a}^\dagger\jop_{\nu,a} \rho - \rho\, \jop_{\nu,a}^\dagger\jop_{\nu,a} \right).
\end{align}
These operators are generically superpositions of lowering operators with coefficients found as the eigenstates of the dissipative interaction matrix $\mathbf{\Gamma}^{a}$ with elements $\Gamma_{jl}^{a}$, with their rates, $\set{\Gamma_\nu^{a}}$, given by the corresponding eigenvalues.

Throughout the paper, we consider that the transition frequencies are all sufficiently distinct such that photons associated with one transition cannot excite any others, and interactions of the form $\hats_{eg_a}^j \hats_{g_be}^l$ are heavily detuned and can be ignored. This condition is naturally met for transitions from an excited state to states with different angular momentum. For transitions to different Zeeman levels, we assume the presence of a magnetic field to break the degeneracy. This requires the Zeeman splitting to be much larger than the linewidth of the emitted light. The spectrum is maximally broadened for two-level atoms at the same spatial location, as this situation produces the shortest possible burst. In this case, one requires a magnetic field such that $B \gg N\hbar \Gamma_0 / \mu_B$, where $\Gamma_0$ is the bare decay rate of a single atom. This corresponds to magnetic fields on the order of $\sim 100$G for the atom numbers considered here. For 2D arrays, the power spectrum is expected to scale sublinearly with the atom number, thus requiring smaller Zeeman shifts.

Throughout the paper, we plot the emission rate of photons by such arrays. The instantaneous total emission rate from the array is directly proportional to the derivative of the population and is calculated as
\begin{equation}
R = -\frac{\mathrm{d}}{\mathrm{d}t}\sum\limits_{j=1}^N\braket{\hat{\sigma}_{ee}^j} = \sum\limits_{\nu=1}^N\sum\limits_{a} \Gamma_\nu^a \braket{\jop_{\nu,a}^\dagger\jop_{\nu,a}}.
\end{equation}
The emission rate of light on a particular transition $\ket{e}\rightarrow\ket{g_a}$ is instead
\begin{equation}
R_a = \frac{\mathrm{d}}{\mathrm{d}t}\sum\limits_{j=1}^N\braket{\hat{\sigma}_{g_ag_a}^j} = \sum\limits_{\nu=1}^N \Gamma_{\nu}^a \braket{\jop_{\nu,a}^\dagger\jop_{\nu,a}}.
\end{equation}
If light is only measured in a particular direction, we make use of the directional collective operator~\cite{Carmichael00,Clemens03}
\begin{align}
\mathcal{\dop}_{a}(\theta,\varphi) = &\sqrt{\frac{3\Gamma_0^{a}}{8\pi} \left[ 1 - \left(\db_{a}\cdot\mathbf{\mathcal{R}}(\theta,\varphi) \right)^2 \right] \mathrm{d}\Omega} \notag \\ &\;\;\;\;\;\;\;\;\;\;\;\;\;\;\;\;\times\sum\limits_{j=1}^N \mathrm{e}^{-\ii k_0^{a} \mathbf{\mathcal{R}}(\theta,\varphi)\cdot\rb_j} \hats_{g_ae}^j,
\end{align}
to find 
\begin{align}
R_a(\theta,\varphi) &= \braket{\dop^\dagger_a(\theta,\varphi)\dop_a(
\theta,\varphi)}\notag \\&\propto \Gamma_0^a\sum\limits_{j,l=1}^N \mathrm{e^{\ii k_0^a \mathbf{\mathcal{R}}(\theta,\varphi)\cdot(\rb_l-\rb_j)}} \braket{\hat{\sigma}_{eg_a}^l\hat{\sigma}_{g_a e}^j}.
\end{align}
In these expressions, $k_0^a$ is the wave vector of the transition $\ket{e}\rightarrow\ket{g_a}$, $\mathbf{\mathcal{R}}(\theta,\varphi)$ is a unit vector in a direction $(\theta,\varphi)$, and $\mathrm{d}\Omega$ is a solid-angle increment.

\subsection{Conditions for many-body superradiance}

\subsubsection{Two-level systems}

The emission of a superradiant burst can be predicted using the set of eigenvalues of the dissipative interaction matrix $\mathbf{\Gamma}$, with elements $\Gamma_{jl}$, $\set{\Gamma_\nu}$~\cite{Masson22}. The minimal requirement for a superradiant burst is an initial positive slope in the emitted photon rate or, equivalently, that the emission of the first photon \emph{on average} enhances the emission rate of the second. In previous work, we have shown that the necessary criterion for a superradiant burst to be emitted from an initially fully excited ensemble of $N$ two-level atoms is~\cite{Masson22}
\begin{equation}
\mathrm{Var.}\left(\frac{\set{\Gamma_\nu}}{\Gamma_0}\right) \equiv \frac{1}{N} \sum\limits_{\nu=1}^N \left( \frac{\Gamma_\nu^2}{\Gamma_0^2} - 1 \right) > 1.\label{g2basic}
\end{equation}

\subsubsection{General case in which not all light is collected}

The previous equation is derived under the assumption that all emitted light is collected. If that is not the case, one instead has to consider the rate of emission into the optical modes that are detected. The above condition states that the emission of a photon at $t=0$ enhances the emission of a second one. If only a fraction of the light is collected, a superradiant burst is measured only if the first photon enhances the emission of a second \textit{into the optical modes that are detected}. This is calculated as
\begin{equation}
    \tilde{g}^{(2)}(0) \equiv \frac{\sum\limits_{d,a}\Gamma_a\Gamma_d\braket{ \mathcal{O}_a^\dagger \mathcal{O}_d^\dagger \mathcal{O}_d \mathcal{O}_a}}{\left(\Gamma_d\sum\limits_d \braket{\mathcal{O}_d^\dagger\mathcal{O}_d}\right)\left(\Gamma_a\sum\limits_a\braket{\mathcal{O}_a^\dagger\mathcal{O}_a}\right)},\label{eq:mostgenericg2}
\end{equation}
where $\set{\mathcal{O}_d}$ is the set of operators that produces detected photons at rates $\Gamma_d$ and $\set{\mathcal{O}_a}$ is the complete set of operators that produces photons at rates $\Gamma_a$. In the following, we treat two specific cases: emission on multiple transitions and directional decay. Further detail on these derived bounds are given in Appendix A.

\subsubsection{Decay to multiple ground states}

If there are multiple ground states, a superradiant burst is emitted by the fully excited state during decay to $\ket{g_a}$ if
\begin{equation}
\mathrm{Var.}\left(\frac{\set{\Gamma_{\nu}^{a}}}{\Gamma_0^{a}}\right) > \frac{\Gamma_0}{\Gamma_0^{a}},\label{g2transition}
\end{equation}
where $\Gamma_0 = \sum_{a}\Gamma_0^{a}$ is the total decay from the excited state. This is of the same form as Eq.~\eqref{g2basic}, but the enhancement provided by the operators on the particular channel needs to additionally overcome competition between different ``internal'' channels. If all atoms are at a point, then the condition for a superradiant burst on a particular transition reduces to $\Gamma_0^{a}/\Gamma_0 > 1/ (N-1)$.

\subsubsection{Directional decay}

In experiments, light is typically only collected in a particular direction. Directional superradiance is defined as the rate of photon emission into a particular direction having a positive slope, and can persist to much larger interatomic separations than when the entire emitted field is considered~\cite{Robicheaux21PRA_Directional}. As shown in Appendix A, directional superradiance occurs if
\begin{equation}
\sum\limits_{j,l = 1}^N \mathrm{e}^{\ii k^{a}_0\mathbf{\mathcal{R}}(\theta,\varphi)\cdot\left(\rb_l - \rb_j\right)} \frac{\Gamma^{a}_{jl}}{N\Gamma^{a}_0} > 1 + \frac{\Gamma_0}{\Gamma_0^{a}}\label{g2dircondition}.
\end{equation}
Here, we map directional photon detection to atomic emission, where $\mathbf{\mathcal{R}}(\theta,\varphi)$ is a unit vector in the direction of the detector and $k_0^{a} = 2\pi / \lambda_0^{a}$ the wave vector of the transition~\cite{Carmichael00,Clemens03}. We define the quantity
\begin{equation}
S = \frac{\sum\limits_{j,l = 1}^N \mathrm{e}^{\ii k^{a}_0\mathbf{\mathcal{R}}(\theta,\varphi)\cdot\left(\rb_l - \rb_j\right)} \Gamma^a_{jl}}{N\left(\Gamma_0^{a} + \Gamma_0\right)},\label{plottedineq}
\end{equation}
such that if $S < 1$ the photon emission rate will decay monotonically in time, and if $S>1$, the minimal conditions for superradiance are met.

\subsection{Master-equation evolution by cumulant expansion}

It would be ideal to calculate the full dynamics to verify our predictions but the full Hilbert space scales exponentially with the atom number. We approximate the full dynamics by means of a second-order cumulant expansion~\cite{Kramer15,Robicheaux21PRA_HigherOrderMeanField,Rubies23PRR}. This method involves truncating the hierarchy of operator expectation values such that
\begin{equation}
\braket{\hat{u}\hat{v}\hat{w}} = \braket{\hat{u}\hat{v}}\braket{\hat{w}} + \braket{\hat{v}\hat{w}}\braket{\hat{u}} + \braket{\hat{u}\hat{w}}\braket{\hat{v}} - 2\braket{\hat{u}}\braket{\hat{v}}\braket{\hat{w}}.
\end{equation}
The complexity of this expansion scales as $\mathcal{O}(N^3)$ rather than exponentially. Further details are provided in Appendix~B. The accuracy of this approximation is not well characterized for 2D arrays. We benchmark the method in Appendix~C, showing that, generically, the accuracy is better for larger lattice constants.

\begin{figure}[b!]
\includegraphics[width=0.45\textwidth]{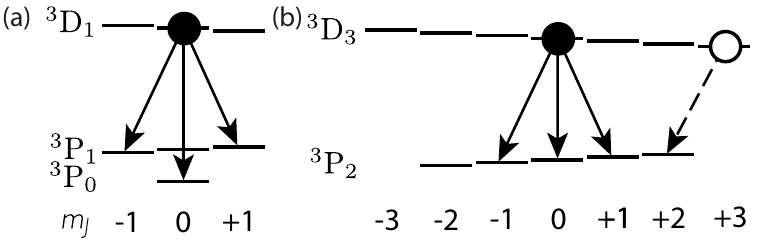}
\caption{The considered level structures. Atoms are initialized in the (a) $^{3}$D$_{1}\ket{J=1,m_J=0}$ or (b) $^{3}$D$_{3}\ket{J=3,m_J=0,3}$ state. In both $m_J=0$ cases, the internal structure is simplified into a toy model with three decay channels: a dominant linear $\pi$-polarized channel and two circularly polarized channels. In (a), the $^{3}$P$_1\ket{J=1,m_J=0}$ state is not considered, as the transition is forbidden.\label{Fig4}}
\end{figure}

\begin{figure*}[t!]
\includegraphics[width=0.9\textwidth]{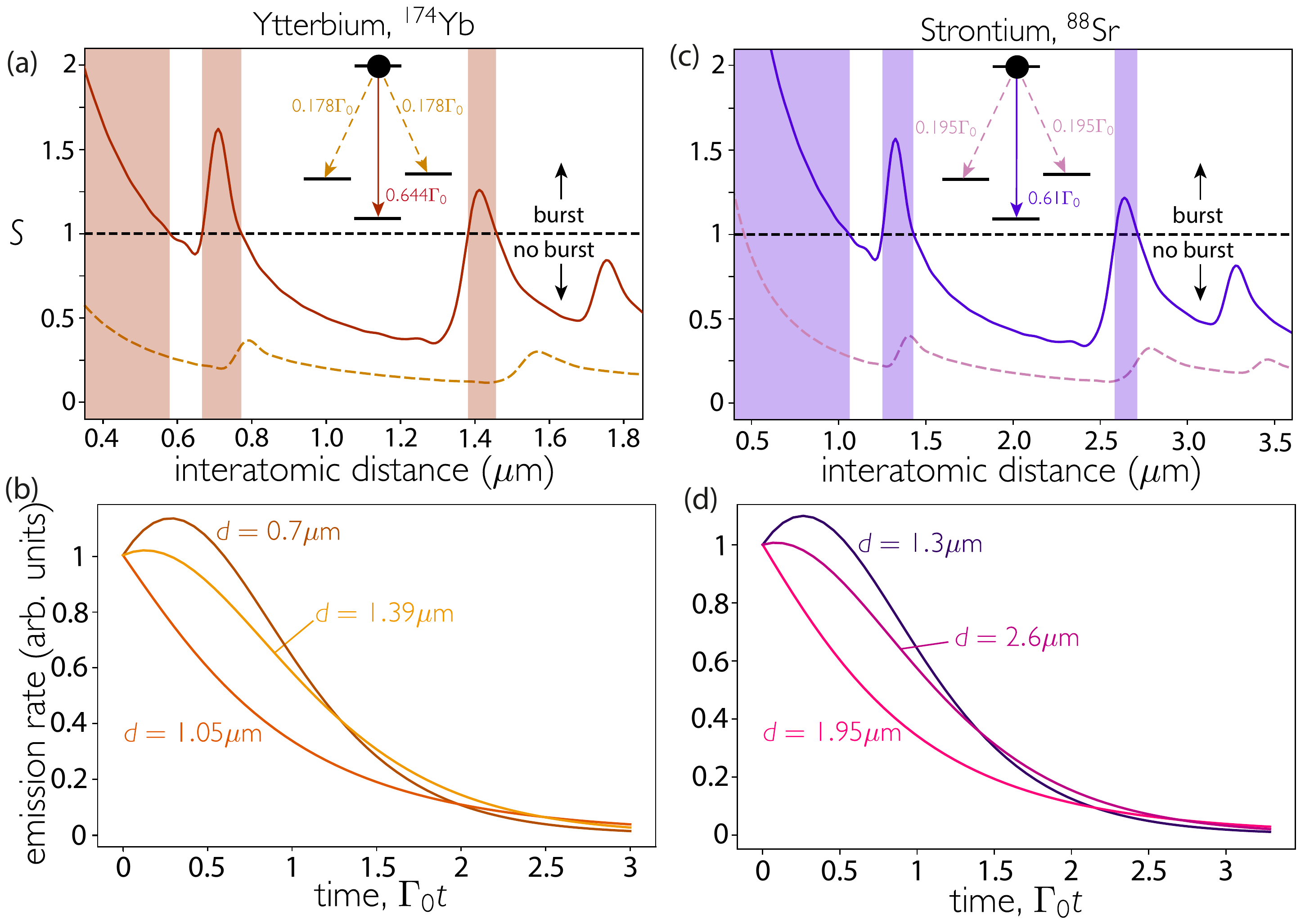}
\caption{Predictions of a burst versus lattice constant in $12\times12$ arrays of (a),(b) $^{174}$Yb and (c),(d) $^{88}$Sr. The atoms are prepared in the $^3$D$_1\ket{J=1,m_J=0}$ state initially and allowed to decay, with the linear transition polarized perpendicular to the array. Light is detected along the $x$-axis. (a,c)~The shaded areas indicate a burst on the $^3$D$_1\ket{J=1,m_J=0} \rightarrow ^3$P$_0\ket{J=0,m_J=0}$ transition (red line), as the quantity $S$ in Eq.~\eqref{plottedineq} is larger than 1 (black dashed line). (b,d) Approximations of the full dynamics via a second-order cumulant expansion (on the $^3$D$_1 \rightarrow ^3$P$_0$ transition) at three particular distances. As predicted by the superradiance condition, as the distance increases, the superradiant burst disappears and then reappears.\label{Fig5}}
\end{figure*}

\section{Superradiance in 2D arrays of alkaline-earth atoms\label{results}}

We now consider AEA arrays in which atoms are initialized in either one of the $^{3}$D$_{1}\ket{J=1,m_J=0}$ or $^{3}$D$_{3}\ket{J=3,m_J=\{0,3\}}$ states and allowed to decay [see Fig.~\ref{Fig4}]. Large inversion can be achieved with a short intense pulse of duration $\tau \ll \left(N\Gamma_0\right)^{-1}$ to prevent collective effects~\cite{Ma23}.

Decay from $^{3}$D$_{1}\ket{J=1,m_J=0}$ can be simplified using information from Tables~\ref{Table1} and ~\ref{Table2}. First, decay to $^3$P$_2$ has minimal impact on the dynamics due to the reduced linewidth. Second, the subsequent decay from $^3\mathrm{P}_1\,\rightarrow\,^1\mathrm{S}_0$ will not impact the initial burst as the decay is not fast enough; nor, due to the short wavelength, will it be strongly collectively enhanced. This leads to a four-level system, as shown in Fig.~\ref{Fig4}(a), with a bright linearly polarized decay channel $^{3}$D$_{1}\ket{J=1,m_J=0}~\rightarrow~^{3}$P$_{0}\ket{J=0,m_J=0}$ and two dimmer circularly polarized transitions $^{3}$D$_{1}\ket{J=1,m_J=0}~\rightarrow~^{3}$P$_{1}\ket{J=1,m_J=\pm1}$. Note that the Clebsch-Gordan coefficient is zero for the $^{3}$D$_{1}\ket{J=1,m_J=0}~\rightarrow~^3$P$_1\ket{J=1,m_J=0}$ pathway. For simplicity, we treat decay to $^{3}$P$_{1}\ket{J=1,m_J=\pm1}$ as split by large enough Zeeman shifts that photons on each transition are not seen by the other, but not by enough to significantly alter the wavelength of the transitions. Without such Zeeman shifts, photons of one circular polarization can drive transitions with the other, allowing the atoms to explore the full Zeeman structure and adding far greater complexity to the problem~\cite{Asenjo19}. A similar structure of three decay channels would be obtained for initialization in $^{3}$D$_1\ket{J=1,m_J = \pm 1}$, but here the brightest transition is circularly polarized which is generically less favorable than linearly polarized transitions for superradiance~\cite{Sierra22}.

We also consider atoms initialized in the $^{3}$D$_{3}\ket{J=3,m_J=0}$ state. From here, there are also three decay channels, as shown in Fig.~\ref{Fig4}(b). As before, the brightest decay channel is linearly polarized, that to $^{3}$P$_{2}\ket{J=2,m_J=0}$, and the two dimmer decay channels to $^{3}$P$_{2}\ket{J=2,m_J=\pm1}$ are circularly polarized. As above, we assume that these channels are independent due to sufficiently large Zeeman shifts. Alternatively, atoms initialized in $^{3}$D$_{3}\ket{J=3,m_J=3}$ are two-level systems with circularly polarized decay, as the only decay channel is to $^{3}$P$_{2}\ket{J=2,m_J=2}$. To study this situation, we consider a rotated magnetic field such that the dipole moment of the two-level systems is oriented as $\db=\sqrt{1/2}\left( \hat{y} + \ii\hat{z}\right)$, so that the detector position is still perpendicular to the polarization axis, enhancing the signal. Other three- (and two-) decay-channel systems could also be obtained by starting in different Zeeman levels in the $^3$D$_3$ line.

We thus reduce the level structure of both atomic species to those shown in Fig.~\ref{Fig4}. Starting from states with $m_J=0$, the master equation in Eq.~\eqref{fullme} reduces to
\begin{equation}
\dot{\rho} = -\frac{\ii}{\hbar} [\mathcal{H}_{f} + \mathcal{H}_{g} + \mathcal{H}_{h},\rho] + \mathcal{L}_{f}(\rho) + \mathcal{L}_{g}(\rho) + \mathcal{L}_{h}(\rho), \label{reducedme}
\end{equation}
where $\ket{e}$ is the excited state and $\ket{f,g,h}$ are the three ground states. For the two-level system we instead have
\begin{equation}
\dot{\rho} = -\frac{\ii}{\hbar} [\mathcal{H}_{g},\rho] + \mathcal{L}_{g}(\rho).
\end{equation}

\subsection{Many-body decay versus distance}

We first investigate atoms initialized in the $^{3}$D$_{1}\ket{J=1,m_J=0}$ state. We consider the condition given in Eq.~\eqref{g2dircondition} for the specific case of a square array of $12\times12$ atoms. The detector is placed along the $x$-axis, which should see significant emission as it is perpendicular to the dipole moment. For $^{174}$Yb atoms, this detector would measure a superradiant burst on the dominant transition for any interatomic separation satisfying $d < 0.6\,\mu$m, as shown in Fig.~\ref{Fig5}(a). This distance would be challenging for tweezer-array experiments, but is achievable in an optical lattice~\cite{Rui20,Srakaew23}.

Superradiance can also be observed at particular ``islands'' where the set of decay operators combines to realize a sudden revival in emission in a particular direction. For this detector position, this occurs in regions centered on $d=0.7\,\mu$m and $d=1.4\,\mu$m, corresponding to a half and full wavelength of the brightest transition respectively. These revivals are due to geometric resonances, where a mode that emits in this direction suddenly increases in amplitude due to Umklapp scattering, and thus becomes significantly brighter~\cite{Nienhuis87,Bettles16PRA}. While these processes can also revive global superradiance~\cite{Masson22,Sierra22}, the effect is much more pronounced for directional superradiance~\cite{Robicheaux21PRA_Directional}. Indeed, at these distances no superradiant burst would be observed if all light was collected. Furthermore, this type of superradiance is highly dependent on the detector position, so the predicted distances change as a function of the detector angle. This is explored in more detail in Appendix D. The approximated full dynamics (via second-order cumulant expansion) agree with our predictions, as shown in Fig.~\ref{Fig5}(b).

Superradiance can also be observed in $^{88}$Sr. The dominant decay channel from $^{3}$D$_{1}\ket{J=1,m_J=0}$ has a smaller branching ratio than that in $^{174}$Yb, but due to its much longer wavelength, the constraints on the interatomic distance are less tight. Figure~\ref{Fig5}(c) shows that a superradiant burst is always observed for $d<1\,\mu$m. In addition, superradiance could be observed at the revivals with interatomic spacing $1.3\,\mu$m and $2.6\,\mu$m. Therefore, as in $^{174}$Yb, as the interatomic spacing is increased, directional superradiance disappears and reappears. Strontium thus also offers a suitable platform for the direct observation of Dicke superradiance, despite the less favorable branching ratios to each state (see Tables~\ref{Table1} and \ref{Table2}).

\begin{figure}[b!]
\begin{center}
\includegraphics[width=0.495\textwidth]{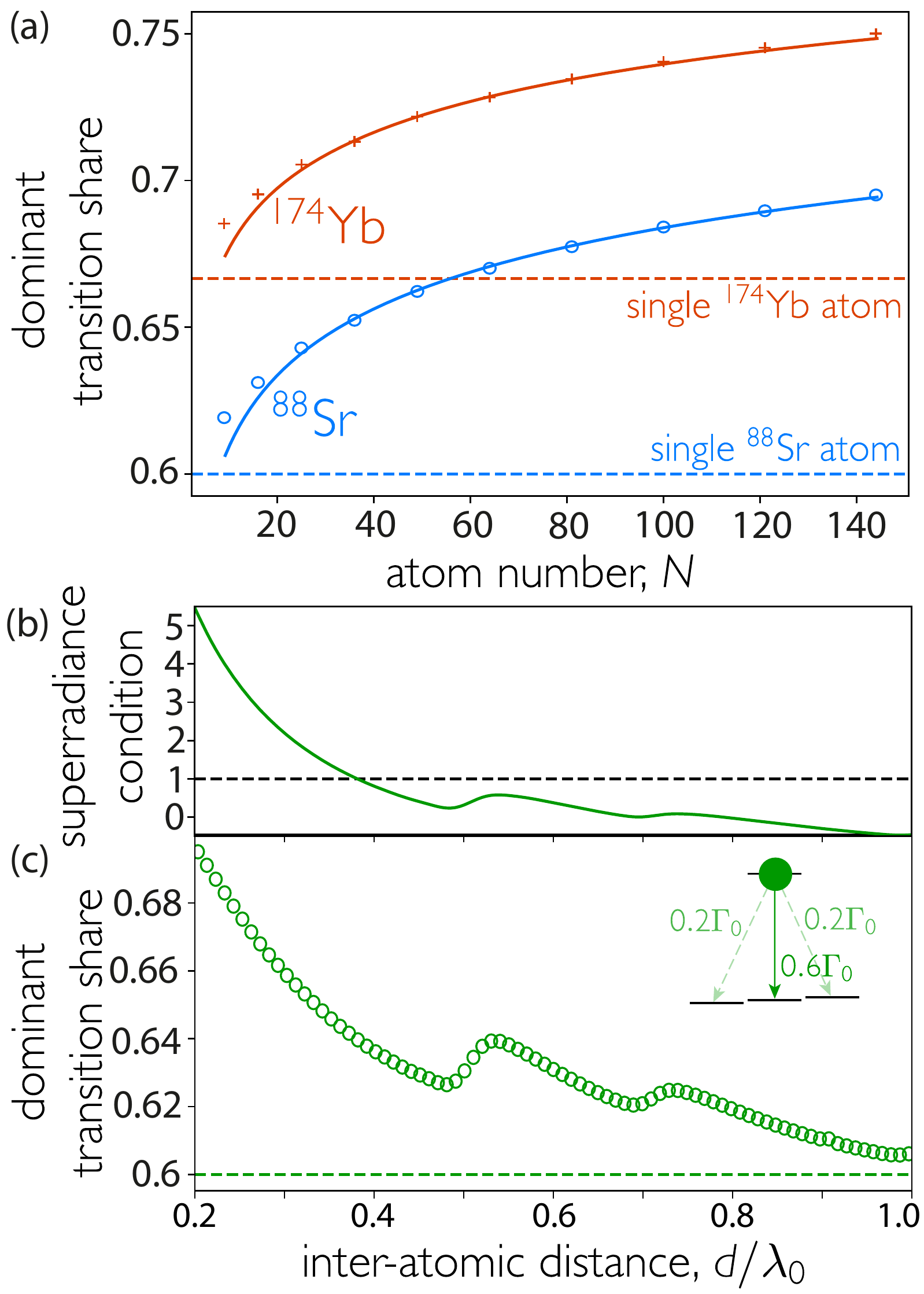}
\caption{The manipulation of branching ratios by collective emission. (a) The scaling of the total share of light emitted on the brightest linearly polarized transition (polarized perpendicular to the array) with the atom number for square arrays of spacing $d = 0.2\lambda_0$, obtained by second-order cumulant-expansion simulations. Simulations for $^{174}$Yb ($^{88}$Sr) are plotted in red (blue), with atoms initialized in the $^3$D$_1 \ket{J=1,m_J=0}$ ($^3$D$_3 \ket{J=3,m_J=0}$) state. The horizontal dashed lines represent the branching ratio for independent atoms. The solid lines are best fits to data from $N\geq25$ of the form $1-A/N^B$ with $A=0.400~(0.481)$ and $B=0.093~(0.091)$ for $^{174}$Yb ($^{88}$Sr)} (b,c) $12\times12$ atoms are initialized in the $^3$D$_3 \ket{J=3,m_J=0}$ state. (b) The condition given by Eq.~\eqref{g2transition} in the form $\mathrm{Var.}\left(\set{\Gamma_{\nu}^{cd}}/\Gamma_0^{cd}\right) - \Gamma_0/\Gamma_0^{cd} + 1$; a global superradiant burst will be measured on the transition to $^3$P$_2\ket{J=2,m_J=0}$  where the solid line is above the dashed line. (c)~The total share of the light emitted on the brightest linearly polarized transition as the interatomic distance is changed.\label{Fig6}
\end{center}
\end{figure}

\begin{figure*}[t!]
\includegraphics[width=\textwidth]{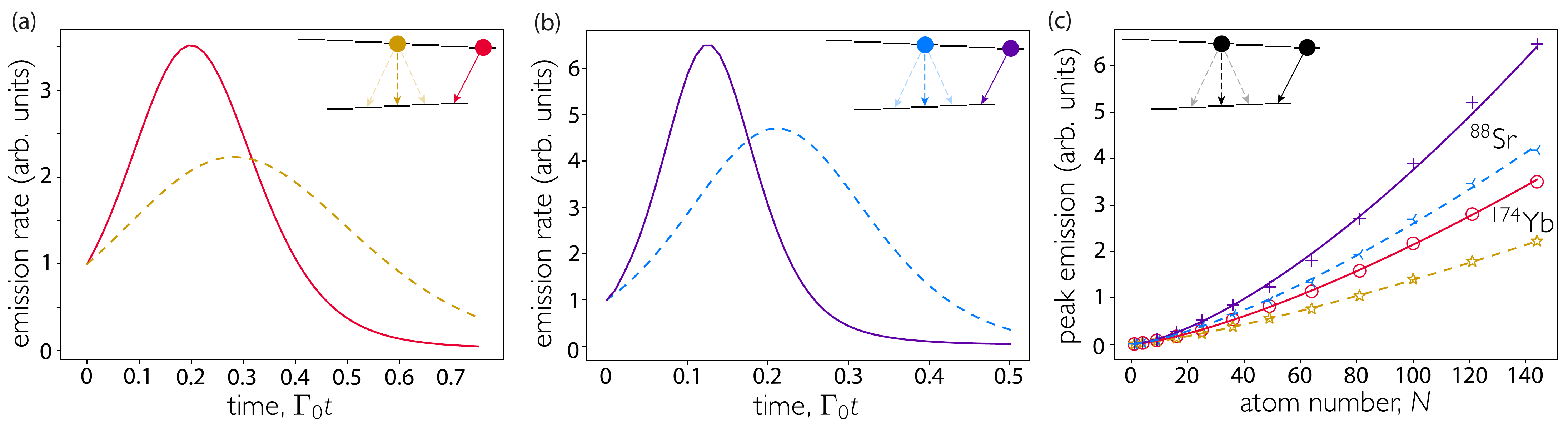}
\caption{The largest superradiant burst in ordered 2D arrays of AEAs. (a) $^{174}$Yb and (b) $^{88}$Sr atoms are initialized in $^3$D$_3 \ket{J=3,m_J=0}$, with the dominant linear transition polarized perpendicular to the array, or $^3$D$_3 \ket{J=3,m_J=3}$, with the only possible decay having $\sqrt{1/2}\left(\hat{y}+\ii\hat{z}\right)$ polarization. Second-order cumulant-expansion simulations of the superradiant burst emitted on the (dashed) $^3$D$_3 \ket{J=3,m_J=0} \rightarrow ^3$P$_2 \ket{J=2,m_J=0}$, with parasitic decays to $\ket{J=2,m_J=\pm 1}$, and (solid) $^3$D$_3 \ket{J=3,m_J=3} \rightarrow ^3$P$_2 \ket{J=2,m_J=2}$ transition. (c) The scaling of the peak emission rate with atom number. The lines are power-law fits of data with $N\geq25$. For the  two-level system, the scaling of the peak is $N^{1.38}$ for $^{174}$Yb and $N^{1.47}$ for $^{88}$Sr, while the four-level system  scales as $N^{1.29}$ for $^{174}$Yb and $N^{1.37}$ for $^{88}$Sr. In all plots, the lattice constant is $d=244$\,nm.\label{Fig7}}
\end{figure*}

\subsection{Collective ``closing'' of the atomic transition}

Once a transition starts to decay superradiantly, it proceeds quickly, ``stealing'' photons from the other transition (as has been discussed for the toy model described in Sec.~\ref{lambda}). The effect is stronger for smaller interatomic distances, because the superradiant burst is much faster in that regime. The order of the lattice structure limits the impact of coherent interactions even at small interatomic distances, in contrast to disordered systems~\cite{Sutherland17}. In Fig.~\ref{Fig6}, we plot the total photon share scattered on the dominant transition during superradiance. For $^{88}$Sr, starting in the $^3$D$_3\ket{J=3,m_J=0}$ state, a single atom scatters $60\%$ of the light on its brightest transition. By contrast, a $12\times12$ array overcomes the Clebsch-Gordan coefficient and scatters almost $70\%$ on that transition. We see a similar improvement at telecom wavelengths for $^{174}$Yb initialized in $^3$D$_1\ket{J=1,m_J=0}$. As for the idealized case of atoms at a point, the light share can be described well by a power law with the atom number.

The geometry of the array dictates the relative scattering on each channel and how much one can go beyond the ratio dictated by the single-atom Clebsch-Gordan coefficients. As the interatomic distance is increased, generally the transition is reopened, as shown in Fig.~\ref{Fig6}(c). However, revivals due to the geometric resonances can be seen by comparison to the variance of the set of decay rates [see Fig.~\ref{Fig6}(b)]. The revivals appear to be relatively minor but are capable of strongly impacting the decay dynamics. It should be noted that the condition for global superradiance is not met in the arrays of larger lattice constants, yet the closing of these weaker channels still occurs despite the fact that the avalanche is relatively weak. As the superradiant burst allows one to overcome the intrinsic branching ratios of decay from a state, many-body superradiance could also be useful to enhance the emission, and thus the collection efficiency, into a given direction or mode.

The share could be further increased by intentionally seeding the transition with a small fraction of the atoms deterministically placed in the desired ground state, or an incomplete initial excitation from a particular ground state. Instead of relying on a quantum fluctuation to drive the start of the superradiant burst, these atoms would provide an artificial fluctuation. This would accelerate the superradiant burst on that seeded transition and generate a large atomic population in that state~\cite{PineiroOrioli22,Asaoka22,Kersten23}. Nevertheless, how effectively this fluctuation will trigger the avalanche depends on its specific spatial profile and phase. We will thus not explore this avenue here.

\subsection{Scaling of the burst}

The largest possible burst occurs for atoms initialized in the $^3$D$_3\ket{J=3}$ manifold. The transition wavelength is $\lambda_0 = 1.80~\,\mu$m for $^{174}$Yb and $\lambda_0=2.92~\,\mu$m in $^{88}$Sr. To minimize the interatomic distance, both species are assumed to be trapped in an optical lattice with 244~nm lattice spacing, corresponding to a wavelength of 488~nm, for which Yb and Sr are trapped in the relevant states and high-power lasers are available. This yields an interatomic spacing of $d = 0.136\lambda_0$ for $^{174}$Yb and $d=0.084\lambda_0$ for $^{88}$Sr.  We consider two initial states: $\ket{J=3,m_J=0}$ and $\ket{J=3,m_J=3}$. In the former case, the decay is to three states, with a dominant transition that is linearly polarized. In the latter case, the atoms become two-level systems, decaying by circular $\sigma^+$-polarized light. This closed two-level transition can be accessed in all AEAs, including fermionic isotopes.

The largest possible burst is emitted by the simplest system operating at the longest wavelength, as shown in Fig.~\ref{Fig7}. For the two-level system, the peak emission rate is more than 3 times greater than the emission rate from an array of $12\times12$ $^{174}$Yb atoms, and more than six times greater for $^{88}$Sr. The peak scales as $\sim N^{1.38}$ and $\sim N^{1.47}$ for $^{174}$Yb and $^{88}$Sr, respectively. The smaller peak emitted from the $\ket{J=3,m_J=0}$ state is still significant, and scales as $\sim N^{1.29}$ for $^{174}$Yb and $\sim N^{1.37}$ for $^{88}$Sr. The exponent in the power law is dependent on the distance between atoms, generically decreasing as the interatomic distance is increased (further details are provided in Appendix E). A word of caution is needed though, as the validity of the second-order cumulant expansion is not well characterized for these large atom numbers at such small distances and may overestimate the burst due to significant multipartite correlations~\cite{Rubies23PRR}.

\section{Conclusions}

We have presented results on collective decay in realistic arrays of alkaline-earth atoms. Building on previous work, we have calculated conditional correlation functions to predict the nature of the collective decay. We have predicted highly nontrivial many-body decay through control of the interatomic spacing of the array and position of the detector. Focusing on the particular cases of $^{88}$Sr and $^{174}$Yb, we have shown that the observation of Dicke superradiance should be feasible in such systems. Furthermore, we show that by increasing the interatomic separation, Dicke superradiance is attenuated and lost, but is then revived at a larger distance. We have shown that this understanding can be used to manipulate how much population ends up in each possible ground state.

Experiments are critical to understand many-body decay, as the full dynamics are only obtained via approximations. We have focused on strontium and ytterbium due to the recent progress in implementing atomic arrays with these species and the favorable set of transitions to achieve subwavelength interatomic spacing. However, similar results should also be possible with other alkaline-earth elements, which have the same structure, but where progress in cooling and trapping is less advanced~\cite{Kraft09,De09,Parker12,Kulosa15}. The relative spacing (and order) of levels is different in all these atoms. For example, in radium, there is a two-level linearly polarized transition - as the $^3$D$_1$ state can only decay to $^3$P$_0$ - at a far-infrared wavelength of $\sim16\,\mu$m~\cite{Guest07}. Rare-earth elements also have infrared transitions from the ground state and can be similarly trapped in short wavelengths due to strong blue transitions~\cite{Lu11,Aikawa12,Hemmerling14,Miao14,Bloch23}.

Our results may be relevant in the context of Rydberg quantum simulators~\cite{Labuhn16,Bernien17,Browaeys20}. Excited Rydberg states can decay via a fast short-wavelength transition (and therefore not collectively enhanced) or via much slower but very long-wavelength transitions. Our work implies that the amount of light scattered on these long Rydberg-Rydberg transitions could be significantly enhanced by collective decay~\cite{Wang07,Goldschmidt16,Sutherland17}. Furthermore, an understanding of the collective enhancement of coupling between black body photons and the $^3$P$_0$ state in atomic optical lattice clocks is key to achieving high precision in compact devices~\cite{Beloy12,Nicholson15,Hutson24}.

Control of the atoms is translated into control over the emitted light. For example, initial superposition states will emit superpositions of different pulses, with the potential for generation of macroscopic superposition states of light. In particular, the potential for $^{174}$Yb arrays to produce nonclassical light at telecom frequencies is tantalizing. While we have focused on the interatomic spacing of the array and the relative position of the detector to control the decay, there are additional tuning knobs that could be harnessed. The dynamics - and in particular the directionality - could be altered by changing the geometry of the array, either by modifying the lattice or the global shape. Manipulation of the initial state, adding coherent or incoherent drives, or manual addition of site-specific inhomogeneity~\cite{Rubies22PRR} - all of this will impact the dynamics and steady state of the system. 

An understanding of the various decay processes - and the freezing out of coherent dynamics - opens up possibilities to harness them. For instance, the complex dissipative dynamics provide a method to access highly entangled dark states that completely decouple from the environment. The deterministic production of these states and their potential as resources for quantum computing and metrology~\cite{Paulisch16,Henriet19} remains an exciting open problem.

\section{Acknowledgements}

We are grateful to Francis Robicheaux, Oriol Rubies-Bigorda, Silvia Cardenas-Lopez, Debayan Mitra and Hannes Bernien for useful discussions. We thank Wai-Keong Mok for useful comments regarding the scaling with atom number of the light emitted on each transition. A.A.-G. acknowledges support by the National Science Foundation through the Faculty Early Career Development Program (CAREER) Award (Grant No. 2047380) and QII-TAQS program (Award No. 1936359), and the Air Force Office of Scientific Research through their Young Investigator Prize (Grant No. 21RT0751), as well as by the A. P. Sloan Foundation. A.A.-G. also acknowledges the Flatiron Institute, where some of this work was performed. A.A.-G. and S.J.M. acknowledge additional support by the David and Lucile Packard Foundation. J.P.C. acknowledges support from the NSF Division of Physics (PHY) (Award No. 2112663). S.W. acknowledges support by the National Science Foundation through the QII-TAQS program (Award No. 1936359) and the Alfred P. Sloan Foundation. We acknowledge computing resources from Columbia University's Shared Research Computing Facility project, which is supported by NIH Research Facility Improvement Grant 1G20RR030893-01, and associated funds from the New York State Empire State Development, Division of Science Technology and Innovation (NYSTAR) Contract C090171, both awarded April 15, 2010.

\appendix

\section{Derivation of directional condition for superradiance on a particular channel\label{appA}}

The condition for superradiance to be detected is that the emission of a first photon enhances the rate of emission into the measured channel, meaning that the conditional second-order correlation function is larger than unity. This approach corresponds to a positive slope in the measured emission rate from the fully excited state, as photon emission is the only possible evolution from such a state. We can show this correspondence by noting that the requirement of positive slope in the measured emission rate is the same as the requirement that $R_d(\delta t) > R_d(0)$ as $\delta t \rightarrow 0$. Here, to connect to Eq.~\eqref{eq:mostgenericg2}, we introduce a generic ``detected emission rate'' operator,
\begin{equation}
\Rdet = \sum\limits_{d} \Gamma_d \jop_d^\dagger \jop_d,
\end{equation}
where $\set{\mathcal{O}_d}$ is the set of operators that produces detected photons at rates $\Gamma_d$. For sufficiently small $\delta t$ then $\rho(\delta t) = \rho(0) + \delta t \dot{\rho}(0)$, such that
\begin{equation}
    \braket{\Rdet(\delta t)} = \braket{\Rdet(0)} + \mathrm{Tr.} \left(\Rdet \delta t\dot{\rho}(0)\right).
\end{equation}
This means that for $\braket{\Rdet(\delta t)} > \braket{\Rdet(0)}$, we require $\mathrm{Tr.} \left(\Rdet \delta t\dot{\rho}(0)\right) > 0$ which leads to
\begin{align}
    \sum\limits_{a,d} \Gamma_a &\Gamma_d \braket{\jop^\dagger_a\jop^\dagger_d \jop_d \jop_a} \notag \\&> \sum\limits_{a,d} \frac{\Gamma_a\Gamma_d}{2} \left( \braket{\jop_a^\dagger\jop_a\jop_d^\dagger\jop_d} + \braket{\jop_d^\dagger\jop_d \jop_a^\dagger \jop_a}\right).
\end{align}
Since the fully excited state is an eigenstate of any possible $\jop^\dagger_a \jop_a$, the right-hand side of this expression factorizes and we find our conditional second-order correlation-function condition
\begin{equation}
\sum\limits_{a,d} \Gamma_a \Gamma_d \braket{\jop^\dagger_a\jop^\dagger_d \jop_d \jop_a} > \sum\limits_{a} \Gamma_a \braket{\jop_a^\dagger\jop_a}\sum\limits_{d} \Gamma_d\braket{\jop_d^\dagger\jop_d}.
\end{equation}

\subsection{Multiple ground states}

The derivative of the photon emission rate on a specified transition $\ket{e}\rightarrow\ket{g_a}$ is positive if
\begin{equation}
\frac{\sum\limits_{\mu=1}^N \sum\limits_{b} \sum\limits_{\nu=1}^{N} \Gamma_{\mu}^a\Gamma_{\nu}^b \braket{\jop_{\nu,b}^\dagger\jop_{\mu,a}^\dagger\jop_{\mu,a}\jop_{\nu,b}}}{\left(\sum\limits_b\sum\limits_{\nu=1}^N\Gamma_\nu^b\braket{\jop_{\nu,b}^\dagger\jop_{\nu,b}}\right)\left(\sum\limits_{\mu=1}\Gamma_\mu^a\braket{\jop_{\mu,a}^\dagger\jop_{\mu,a}}\right)} > 1,
\end{equation}
On a fully excited initial state, this expression reads
\begin{align}
1 + \frac{\sum\limits_{\nu=1}^N (\Gamma_v^{a})^2}{N^2\Gamma_0^{a} \Gamma_0} - \frac{1}{N} - \frac{\Gamma_0^{a}}{N\Gamma_0} > 1,
\end{align}
which simplifies to
\begin{align}
\mathrm{Var.}\left(\frac{\Gamma_\nu^{a}}{\Gamma_0^{a}}\right) \equiv \frac{1}{N} \sum\limits_{\nu=1}^N \left[ \left(\frac{\Gamma_\nu^{a}}{\Gamma_0^{a}} \right)^2 - 1\right] > \frac{\Gamma_0}{\Gamma_0^a}
\end{align}
where $\Gamma_0=\sum_a\Gamma_0^a$ is the total excited-state decay rate.

\subsection{Directional decay}

Detection of a photon from a given transition in the far field in a direction governed by spherical angles $\set{\theta,\varphi}$ can be mapped to the collective lowering operator~\cite{Carmichael00,Clemens03},
\begin{align}
\mathcal{\dop}_{a}(\theta,\varphi) = &\sqrt{\frac{3\Gamma_0^{a}}{8\pi} \left[ 1 - \left(\db_{a}\cdot\mathbf{\mathcal{R}}(\theta,\varphi) \right)^2 \right] \mathrm{d}\Omega} \notag \\ &\;\;\;\;\;\;\;\;\;\;\;\;\;\;\;\;\times\sum\limits_{j=1}^N \mathrm{e}^{-\ii k_0^{a} \mathbf{\mathcal{R}}(\theta,\varphi)\cdot\rb_j} \hats_{g_ae}^j,
\end{align}
where $\mathbf{\mathcal{R}}(\theta,\varphi)$ is a unit vector in the direction of the detector, $\mathrm{d}\Omega$ is a solid-angle increment and $k_0^{a}$ is the wave vector of the transition. Using these, the derivative of the photon emission rate on a specified transition $\ket{e}\rightarrow\ket{g_a}$ in a direction $\{\theta,\varphi\}$ is positive if
\begin{equation}
\frac{  \sum\limits_b\sum\limits_{\nu=1}^{N}  \braket{\jop_{\nu,b}^\dagger\dop_{a}^\dagger(\theta,\varphi) \dop_{a}(\theta,\varphi)\jop_{\nu,b}}}{\sum\limits_b\sum\limits_{\nu=1}^N\braket{\jop_{\nu,b}^\dagger\jop_{\nu,b}}\braket{\dop_{a}^\dagger(\theta,\varphi)\dop_{a}(\theta,\varphi)}} > 1.
\end{equation}
On a fully excited state, this expression reads,
\begin{equation}
1 + \frac{\sum\limits_{j,l=1}^N \mathrm{e}^{\ii k_0 \mathbf{\mathcal{R}}(\theta,\varphi)\cdot(\rb_l - \rb_j)} \Gamma_{jl}}{N^2\Gamma_0}  - \frac{1}{N} - \frac{\Gamma_0^{a}}{N\Gamma_0} > 1
\end{equation}
where we have employed that
\begin{equation}
\sum\limits_{\nu=1}^N \Gamma_v^{a} \alpha^*_{\nu,a,j} \alpha_{\nu,a,l} = \Gamma_{jl}^{a}.
\end{equation}
This simplifies to
\begin{equation}
\sum\limits_{j,l = 1}^N \mathrm{e}^{\ii k^{a}_0\mathbf{\mathcal{R}}(\theta,\varphi)\cdot\left(\rb_l - \rb_j\right)} \frac{\Gamma^{a}_{jl}}{N\Gamma^{a}_0} > 1 + \frac{\Gamma_0}{\Gamma_0^{a}}.\label{directionalcondition}
\end{equation}

\section{Second-order cumulant expansion for four-level systems\label{appB}}

We consider four-level atoms that can decay to three ground states $\ket{f,g,h}$ from an excited state $\ket{e}$. Photons associated with each transition $\ket{e}\rightarrow\ket{f,g,h}$ are sufficiently distinct in frequency to not excite one another. To calculate the directional emission rate on a channel $\ket{e} \rightarrow \ket{f}$, we require the evolution of the expectation values of the set $\{\hat{\sigma}_{ef}^i \hat{\sigma}_{fe}^j\}$, meaning that at least a second-order cumulant expansion is necessary. Generically, the evolution of these expectation values depends on the expectation values of sets of three of the population operators, e.g. $\{\hat{\sigma}_{ee}^i\}$ (the fourth can be related to the other three, as the total single-atom population is always unity), all six coherence operators, e.g., $\{\hat{\sigma}_{ef}^i\}$, and all 66 two-operator products, noting that the complex expectation value of the coherence operators leads to extra operators such as $\{\hat{\sigma}_{ef}^i\hat{\sigma}_{fe}^j\}$.

\begin{figure*}[t!]
\begin{center}
\includegraphics[width=0.925\textwidth]{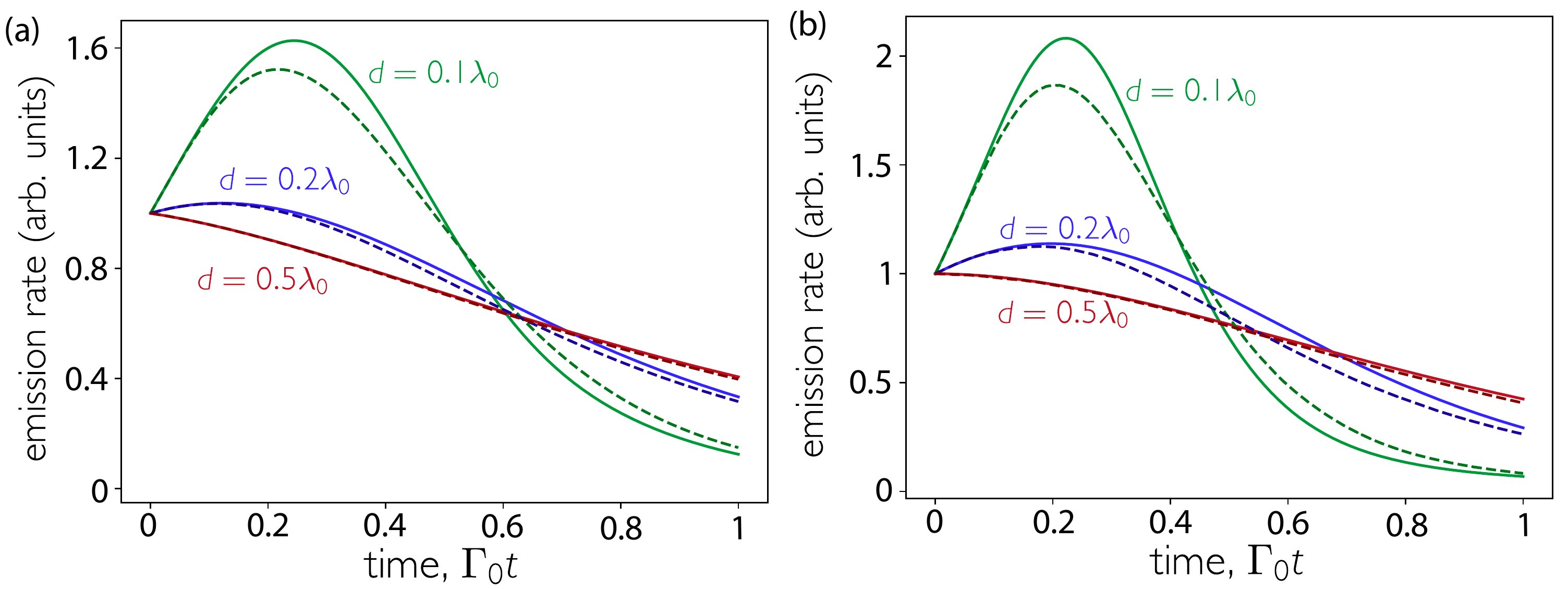}
\caption{The benchmarking of the second-order cumulant expansion against the exact dynamics. Light emitted by a (a) $3\times3$ and (b) $4\times4$ array of two-level systems prepared initially in the excited state, with the polarization axis perpendicular to the array. Light is detected along the $x$-axis. The cumulant-expansion dynamics (solid lines) are compared to exact dynamics (dashed lines), which are obtained from the master equation in (a) and from an ensemble average of 10,000 quantum trajectories in (b).\label{Fig9}}
\end{center}-
\end{figure*}

If the initial state has no coherence, such that initially
\begin{equation}
\{\braket{\hat{\sigma}_{ab}^i}\} = 0 \; \forall \; a\neq b \;\; \mathrm{and} \;\;\; \{\braket{\hat{\sigma}_{ab}^i\hat{\sigma}_{cd}^j}\} = 0 \; \forall \; a \neq b~|~c\neq d,
\end{equation}
then the equations are greatly simplified. This condition is met by the fully excited state or any state where all single-atom states are the ground or excited state. In this case, the single-atom coherences are never different from zero in the second-order cumulant expansion and the expectation values of all two-operator products of the form
$\{\hat{\sigma}_{ab}^i\hat{\sigma}_{ad}^j\}$ are always zero, and those of the form $\hat{\sigma}_{ab}^i \hat{\sigma}_{da}^j$ are zero $\forall \, b \neq d$. The two-operator products of the form $\{\hat{\sigma}_{aa}^i\hat{\sigma}_{bb}^j\}$ and $\{\hat{\sigma}_{ab}^i\hat{\sigma}_{ba}^j\}$ do become nonzero, but only those where one of $a$ or $b$ represents the excited state impact the evolution of the terms needed to calculate the directional emission rate. As such, there is a closed set of operators with expectation values defined as
\begin{subequations}
\begin{align}
a_j &= \braket{\hat{\sigma}_{ee}^j}, b_j= \braket{\hat{\sigma}_{ff}^j}, c_j = \braket{\hat{\sigma}_{gg}^j}, \\
e_{jl} &= \braket{\hat{\sigma}_{ee}^j \hat{\sigma}_{ee}^l}, f_{jl} = \braket{\hat{\sigma}_{ee}^j \hat{\sigma}_{ff}^l}, g_{jl} = \braket{\hat{\sigma}_{ee}^j \hat{\sigma}_{gg}^l}, \\
q_{jl} &= \braket{\hat{\sigma}_{ef}^j \hat{\sigma}_{fe}^l}, p_{jl} = \braket{\hat{\sigma}_{eg}^j \hat{\sigma}_{ge}^l}, r_{jl} = \braket{\hat{\sigma}_{eh}^j \hat{\sigma}_{he}^l}.
\end{align}
\end{subequations}
The expectation values evolve according to
\begin{widetext}
\begin{subequations}\allowdisplaybreaks
\begin{align}
\dot{a}_j &= - \Gamma_0 a_j + \ii \sum\limits_{m\neq j} \left[ - A_{jm}q_{jm} + A_{jm}^*q_{mj} - B_{mj} p_{jm} + B^*_{mj} p_{mj} - C_{jm}r_{jm} + C^*_{mj} r_{mj} \right], \\
\dot{b}_j &= \Gamma_0^{ef} a_j + \ii \sum\limits_{m\neq j}  \left[ - A^*_{mj} q_{mj} + A_{jm} q_{jm}\right], \\
\dot{c}_j &= \Gamma_0^{eg} a_j + \ii \sum\limits_{m\neq j}  \left[ -B^*_{mj}p_{mj} + B_{jm} p_{jm}\right] ,\\
\dot{e}_{jl} &= - 2\Gamma_0e_{jl} + \ii \sum\limits_{m\neq j,l} \left[- A_{jm} \braket{\hat{\sigma}_{ef}^j\hat{\sigma}_{ee}^l\hat{\sigma}_{fe}^m} - A_{lm} \braket{\hat{\sigma}_{ee}^j \hat{\sigma}_{ef}^l \hat{\sigma}_{fe}^m} - B_{jm} \braket{\hat{\sigma}_{eg}^j\hat{\sigma}_{ee}^l\hat{\sigma}_{ge}^m}  - B_{lm} \braket{\hat{\sigma}_{ee}^j \hat{\sigma}_{eg}^l \hat{\sigma}_{ge}^m} - C_{jm} \braket{\hat{\sigma}_{eh}^j\hat{\sigma}_{ee}^l\hat{\sigma}_{he}^m} \right. \notag \\ &\left.- C_{lm} \braket{\hat{\sigma}_{ee}^j\hat{\sigma}_{eh}^l\hat{\sigma}_{he}^m} + A_{mj}^* \braket{\hat{\sigma}_{fe}^j\hat{\sigma}_{ee}^l\hat{\sigma}_{ef}^m} + A_{ml}^* \braket{\hat{\sigma}_{ee}^j\hat{\sigma}_{fe}^l\hat{\sigma}_{ef}^m} + B_{mj}^* \braket{\hat{\sigma}_{ge}^j\hat{\sigma}_{ee}^l\hat{\sigma}_{eg}^m} + B_{ml}^* \braket{\hat{\sigma}_{ee}^j\hat{\sigma}_{ge}^l\hat{\sigma}_{eg}^m} + C_{mj}^* \braket{\hat{\sigma}_{he}^j\hat{\sigma}_{ee}^l\hat{\sigma}_{eh}^m} \right. \notag \\ &\left.+ C_{ml}^* \braket{\hat{\sigma}_{ee}^j\hat{\sigma}_{he}^l \hat{\sigma}_{eh}^m} \right], \\
\dot{f}_{jl} &= - \Gamma_0 f_{jl} - \ii A_{jl} q_{jl} + \ii A^*_{lj} q_{lj} + \Gamma_0^A e_{jl} + \ii \sum\limits_{m\neq j,l} \left[ - A_{jm} \braket{\hat{\sigma}_{ef}^j\hat{\sigma}_{ff}^l\hat{\sigma}_{fe}^m} - A_{ml}^* \braket{\hat{\sigma}_{ee}^j\hat{\sigma}_{fe}^l\hat{\sigma}_{ef}^m} - B_{jm} \braket{\hat{\sigma}_{eg}^j\hat{\sigma}_{ff}^l\hat{\sigma}_{ge}^m} \right. \notag \\ &\left. - C_{jm} \braket{\hat{\sigma}_{eh}^j\hat{\sigma}_{ff}^l\hat{\sigma}_{he}^m} + A_{mj}^* \braket{\hat{\sigma}_{fe}^j\hat{\sigma}_{ff}^l\hat{\sigma}_{ef}^m} + A_{lm} \braket{\hat{\sigma}_{ee}^j\hat{\sigma}_{ef}^l\hat{\sigma}_{fe}^m}+ B_{mj}^* \braket{\hat{\sigma}_{ge}^j \hat{\sigma}_{ff}^l \hat{\sigma}_{eg}^m} + C_{mj}^* \braket{\hat{\sigma}_{he}^j \hat{\sigma}_{ff}^l \hat{\sigma}_{eh}^m} \right],\\
\dot{g}_{jl} &= - \Gamma_0 g_{jl} - \ii B_{jl} p_{jl} + \ii B^*_{lj} p_{lj} + \Gamma_0^B e_{jl} + \ii \sum\limits_{m\neq j,l} \left[ - B_{jm} \braket{\hat{\sigma}_{eg}^j\hat{\sigma}_{gg}^l\hat{\sigma}_{ge}^m} - B_{ml}^* \braket{\hat{\sigma}_{ee}^j\hat{\sigma}_{ge}^l\hat{\sigma}_{eg}^m} - A_{jm} \braket{\hat{\sigma}_{ef}^j\hat{\sigma}_{gg}^l\hat{\sigma}_{fe}^m} \right. \notag \\ &\left. - C_{jm} \braket{\hat{\sigma}_{eh}^j\hat{\sigma}_{gg}^l\hat{\sigma}_{he}^m} + B_{mj}^* \braket{\hat{\sigma}_{ge}^j\hat{\sigma}_{gg}^l\hat{\sigma}_{eg}^m} + B_{lm} \braket{\hat{\sigma}_{ee}^j\hat{\sigma}_{eg}^l\hat{\sigma}_{ge}^m} + A_{mj}^* \braket{\hat{\sigma}_{fe}^j \hat{\sigma}_{gg}^l \hat{\sigma}_{ef}^m} + C_{mj}^* \braket{\hat{\sigma}_{he}^j \hat{\sigma}_{gg}^l \hat{\sigma}_{eh}^m} \right],\\
\dot{q}_{jl} &= -\Gamma_0 q_{jl} - \ii A_{jl} f_{jl} + \ii A_{lj}^* f_{lj} + \Gamma_{lj}^A e_{jl} + \ii \sum\limits_{m\neq j,l}\left[ -A_{mj}^* \braket{\hat{\sigma}_{ee}^j\hat{\sigma}_{fe}^l\hat{\sigma}_{ef}^m}  - A_{lm} \braket{\hat{\sigma}_{ef}^j\hat{\sigma}_{ff}^l\hat{\sigma}_{fe}^m}  + A_{mj}^* \braket{\hat{\sigma}_{ff}^j\hat{\sigma}_{fe}^l\hat{\sigma}_{ef}^m} \right. \notag \\ &\left.+ A_{lm} \braket{\hat{\sigma}_{ef}^j\hat{\sigma}_{ee}^l\hat{\sigma}_{fe}^m}  \right], \\
\dot{p}_{jl} &= -\Gamma_0 p_{jl} - \ii B_{jl} g_{jl} + \ii B_{lj}^* g_{lj} + \Gamma_{lj}^B e_{jl} + \ii \sum\limits_{m\neq j,l}\left[ -B_{mj}^* \braket{\hat{\sigma}_{ee}^j\hat{\sigma}_{ge}^l\hat{\sigma}_{eg}^m} - B_{lm} \braket{\hat{\sigma}_{eg}^j\hat{\sigma}_{gg}^l\hat{\sigma}_{ge}^m}   + B_{mj}^* \braket{\hat{\sigma}_{gg}^j\hat{\sigma}_{ge}^l\hat{\sigma}_{eg}^m} \right. \notag \\ &\left. + B_{lm} \braket{\hat{\sigma}_{eg}^j\hat{\sigma}_{ee}^l\hat{\sigma}_{ge}^m}  \right], \\
\dot{r}_{jl} &= -\Gamma_0 r_{jl} - \ii C_{jl} \left(a_j - e_{jl} - f_{jl} - g_{jl}\right) + \ii C_{jl}^* (a_l - e_{lj} - f_{lj} - g_{lj}) + \Gamma_{ji}^C e_{jl} + \ii \sum\limits_{m\neq j,l}\left[ -C_{mj}^* \braket{\hat{\sigma}_{ee}^j\hat{\sigma}_{he}^l\hat{\sigma}_{ef}^m} \right. \notag \\ &\left.- C_{lm} \braket{\hat{\sigma}_{eh}^j\hat{\sigma}_{hh}^l\hat{\sigma}_{he}^m} + C_{mj}^* \braket{\hat{\sigma}_{hh}^j\hat{\sigma}_{he}^l\hat{\sigma}_{ef}^m} + C_{lm} \braket{\hat{\sigma}_{eh}^j\hat{\sigma}_{ee}^l\hat{\sigma}_{he}^m}  \right] ,
\end{align}
\end{subequations}
\end{widetext}
where we have defined 
\begin{align}
A_{jl} &= J_{jl}^{ef} - \ii \frac{\Gamma_{jl}^{ef}}{2}, B_{jl} = J_{jl}^{eg} - \ii \frac{\Gamma_{jl}^{eg}}{2}, C_{jl} = J_{jl}^{eh} - \ii \frac{\Gamma_{jl}^{eh}}{2},
\end{align}
and three-operator product expectation values are approximated by the second-order cumulant expansion as
\begin{equation}
\braket{\hat{u}\hat{v}\hat{w}} = \braket{\hat{u}\hat{v}}\braket{\hat{w}} + \braket{\hat{v}\hat{w}}\braket{\hat{u}} + \braket{\hat{u}\hat{w}}\braket{\hat{v}} - 2\braket{\hat{u}}\braket{\hat{v}}\braket{\hat{w}}.
\end{equation}

\begin{figure*}[t!]
\includegraphics[width=\textwidth]{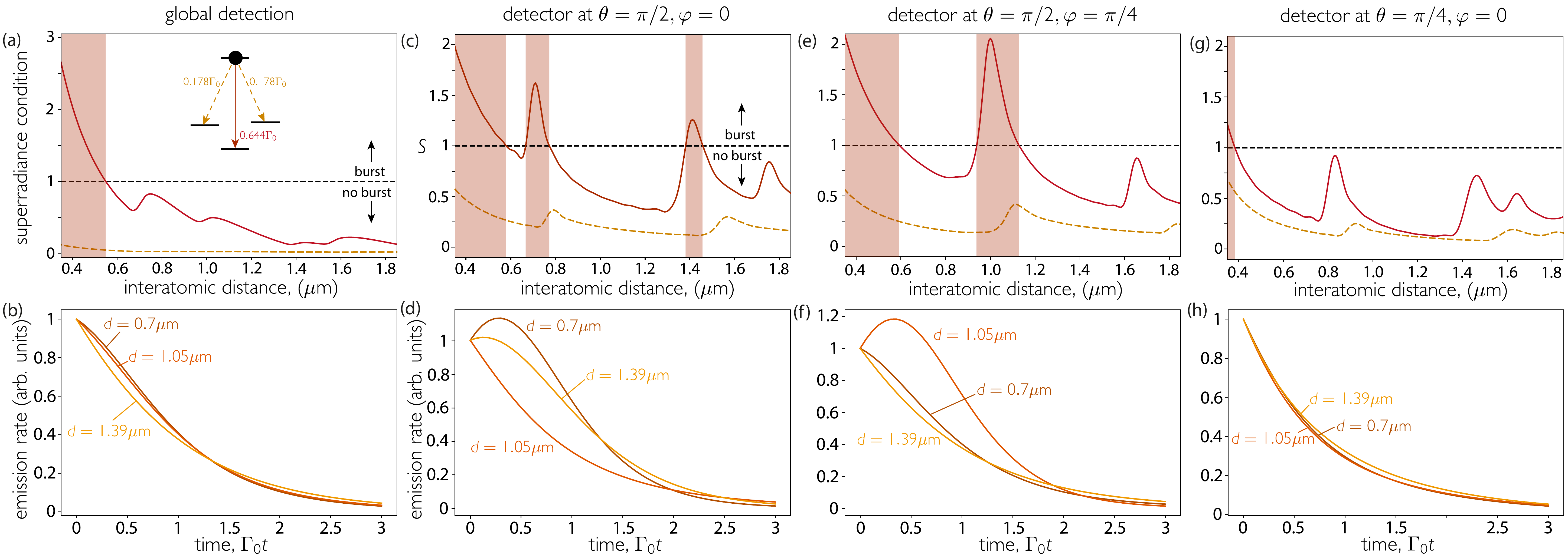}
\caption{Predictions of a burst versus the lattice constant in $12\times12$ arrays of $^{174}$Yb for different detector positions. The atoms are prepared in the $^3$D$_1\ket{J=1,m_J=0}$ state initially and allowed to decay, with the linear transition polarized perpendicular to the array. (a),(b) All light is collected. (c),(d) Light is detected along the $x$-axis. (e),(f) Light is detected in the diagonal of the plane. (g),(h) Light is detected tilted up from the $x$-axis. In (a),(c),(e), and (g), the shaded areas indicate a burst on the $^3$D$_1\ket{J=1,m_J=0} \rightarrow ^3$P$_0\ket{J=0,m_J=0}$ transition (red line). The superradiance condition plotted in (a) is $(\Gamma_0^a/\Gamma_0)\mathrm{Var.}(\set{\Gamma_\nu^a}/\Gamma_0^a)$ and in (c),(e),(g) it is the quantity $S$ given in Eq.~\eqref{plottedineq}. In both cases, a burst is predicted if the condition is larger than 1 (black dashed line). In all cases, the $^3\mathrm{P}_1\ket{J=1,m_J=\pm1}$ transition (orange dashed line) is never superradiant. In (b),(d),(f), and (h) approximations are shown of the full dynamics via a second-order cumulant expansion (on the $^3$D$_1 \rightarrow ^3$P$_0$ transition) at three particular distances. The presence or absence of a superradiant burst is correctly predicted by the superradiance conditions.\label{Fig11}}
\end{figure*}

\section{Benchmarking second-order cumulant expansion}

To benchmark the second-order cumulant expansion, we compare the approximated dynamics to the exact dynamics for small system sizes, as shown in Fig.~\ref{Fig9}. Here, we consider two-level atoms, as calculating the exact dynamics for four-level atoms is not computationally tractable even for 16 atoms. The exact dynamics are found as the ensemble average of quantum trajectories~\cite{Carmichael93}. At short times, and for modest separations, the cumulant expansion is an excellent approximation of the dynamics. As the distance decreases, the error becomes more significant. For $d=0.1\lambda_0$, the peak is overestimated by 12\% for a $4\times4$ array and by 9\% for a $3\times3$ array. The relative error is much larger at later times, as the cumulant expansion is unable to capture the subradiant tail~\cite{Robicheaux21PRA_HigherOrderMeanField}. This is also true for $d=0.2\lambda_0$, where the burst is captured more accurately (overestimated by only 1\% for a $4\times4$ array) but large relative errors occur in the tail.

\section{Directional dependence of emission}

In the main text, we consider light measured by a detector placed along the $x$-axis. We show that the presence or absence of a burst is nontrivially dependent on the distance between atoms. In reality, the placement of the detector defines this nontrivial dependence. Figure~\ref{Fig11} recreates Fig.~\ref{Fig5} but considering different detection schemes.

If all the light were to be collected, a superradiant burst would not be detected for the distances and atom numbers that we consider. As shown in Fig.~\ref{Fig11}(a), a global superradiant burst for the $^3$D$_1\ket{J=1,m_J=0} \rightarrow ^3$P$_0\ket{J=0,m_J=0}$ transition in $^{174}$Yb requires distances of $d\lesssim 0.55\lambda_0$. Fortunately, the conditions for directional superradiance are generically less strict than for global superradiance, allowing for experimental observation at much larger interatomic separations or smaller atom numbers~\cite{Robicheaux21PRA_Directional}.

The measurement of a directional superradiant burst requires correct placement of the detector. As shown in the main text and reproduced here in Fig.~\ref{Fig11}(b,f), if light is detected along the $x$-axis, superradiant bursts could be observed for arrays with an interatomic spacing equivalent to $d=0.5\lambda_0$ and $d=\lambda_0$ but not for $d=0.75\lambda_0$. Conversely, if the detector is rotated by $\pi/4$ degrees in the plane, then a burst could be observed at $d=0.75\lambda_0$ but not at $d=0.5\lambda_0$, nor at $d=\lambda_0$ [see Fig.~\ref{Fig11}(c,g)]. This is because the detector lies along the diagonal axis of the square grid and emission in this direction is enhanced by a geometric resonance at $\lambda_0/\sqrt{2}$.

It is known that disordered samples preferentially emit along the axis of maximum optical depth~\cite{Ernst68,Clemens03}. This holds true in 2D arrays, where emission is preferential in the plane. However, it becomes more complicated due to interference effects producing preference \textit{amongst} the in-plane directions. The effect can be seen by the absence of superradiant bursts sent to a detector positioned significantly out of the plane of the array [see Fig.~\ref{Fig11}(d,h)]. To measure superradiance in such a direction requires even shorter interatomic distances than global superradiance.

\section{Superradiant burst scalings with atom number}

\begin{figure*}
\includegraphics[width=\textwidth]{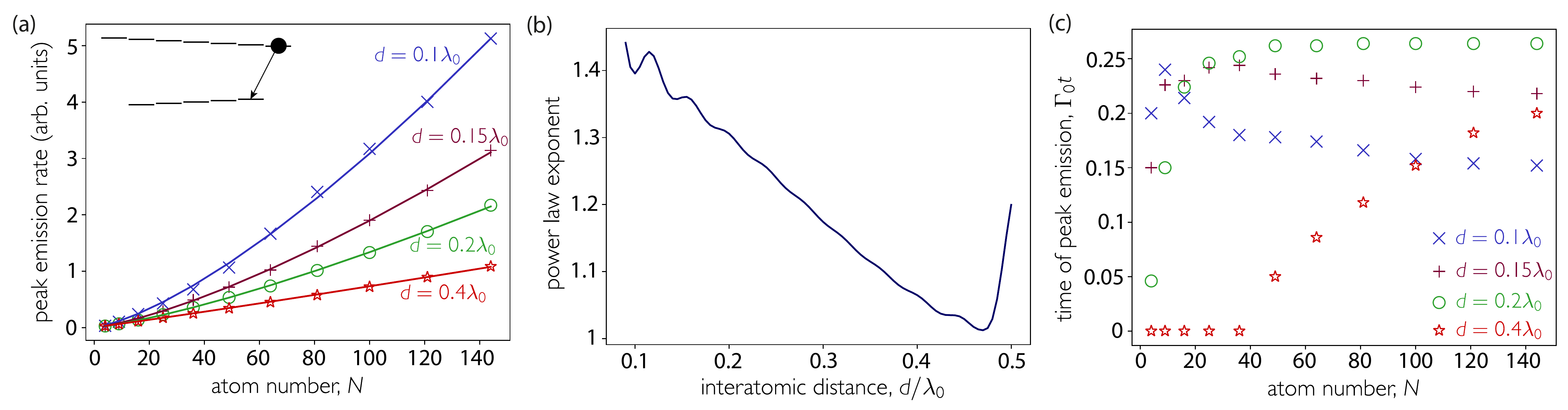}
\caption{Scalings of superradiant bursts in ordered 2D arrays of AEAs. $^{174}$Yb or $^{88}$Sr atoms are initialized in $^3$D$_3 \ket{J=3,m_J=3}$, with the only possible decay having $\sqrt{1/2}\left(\hat{y}+\ii\hat{z}\right)$ polarization. (a) The scaling of the peak emission rate with atom number for different distances. The lines are power-law fits of data for all $N$ such that the peak time occurs at $t>0$. (b) The power-law exponent. (c) The time of the peak emission rate with the atom number for different distances.\label{Fig10}}
\end{figure*}

In the main text, we find the scaling of the superradiant burst with the atom number for a specific distance of $244$~nm. Here, we consider different distances [see Fig.~\ref{Fig10}(a)]. We find that generically, as the distance increases, the power-law exponent reduces. However, as shown in Fig.~\ref{Fig10}(b), the trend is not purely monotonic, featuring weak oscillations. This is most likely due to finite-size effects. As the interatomic distance approaches $0.5\lambda_0$ the exponent suddenly increases again, due to the geometric resonances discussed in the main text. There seems to be a correlation between the size of the burst for a given atom number and the rate of growth as the atom number is increased.

The time of the peak is more complicated, as shown in Fig.~\ref{Fig10}(c). One expects, from Dicke's case of atoms at a point, that the time of the peak should scale as $\mathrm{log}(N)/N$, reducing as the number of atoms increases and the peak grows. This is because the peak occurs as the mean number of excited atoms is $N/2$, which is reached faster as the number of atoms increases. This can be seen for $d=0.1\lambda_0$, where for $N\geq9$ we see a monotonic decrease in the time of the peak emission rate. However, for other distances this is not the pattern, due to a competing factor: induced dephasing. Dephasing causes the peak to occur above half excitation. As the relative level of dephasing decreases, the fraction of excited atoms at the peak emission rate tends toward half. Since increasing the number of atoms makes superradiance overcome dephasing, the time of the peak can increase with the atom number. This is shown clearly for $d=0.4\lambda_0$ in Fig.\ref{Fig10}(c). However, eventually this effect saturates and the time of peak emission rate begins to decrease again, as shown for $d=0.15\lambda_0$.

\end{document}